\documentclass[cup7b]{cupbook}
\usepackage{fancyhdr}
\usepackage{graphicx,calc,epsfig}
\usepackage{amsmath,amssymb,amsfonts}

\def\S{\scriptstyle}

\def\bb1{\textup{\small{1}} \kern-3.6pt \textup{1}\ }

\DeclareFontFamily{U}{rsfs}{}         
\DeclareFontShape{U}{rsfs}{m}{n}{<5> rsfs5 <6><7> rsfs7          %
  <8><9><10><10.95><12><14.4><17.28><20.74><24.88> rsfs10}{}     %
\DeclareMathAlphabet{\mathfs}{U}{rsfs}{m}{n}                     %
\newcommand{\mfs}[1]{\mathfs {#1}}                               %

\newcommand{\sH}{{\mfs H}}

\newcommand{\va}{\scriptscriptstyle}
\newcommand{\vani}{\scriptstyle}
\newcommand{\R}{\mathbb{R}}

\newcommand{\Hp}{{\sH}_{phys}}

\newcommand{\Hk}{{\sH}_{kin}}

\newcommand{\PP}{{ P}}
\newcommand{\So}{{\hat {\cal S}}}

\newcommand{\BC}{Barrett-Crane }

\newcommand{\be}{\nopagebreak[3]\begin{equation}}
\newcommand{\ee}{\end{equation}}
\newcommand{\ba}{\nopagebreak[3]\begin{eqnarray}}
\newcommand{\ea}{\end{eqnarray}}

\usepackage{bbm}

\begin{document}

\newtheorem{theorem}{Theorem}[chapter]
\newcommand{\blackboard}{\bf }

\hyphenation{tele-vision} \hyphenation{Quantum spacetime in LQG}

\pagenumbering{roman}
\maketitle 

\pagenumbering{arabic}

\author[Alejandro Perez]{Alejandro Perez\\ Centre de Physique
Th\'eorique,\\ Campus de Luminy, 13288 Marseille, France.\\ 
 Unit\'e Mixte de Recherche (UMR 6207) du \\
CNRS et des Universit\'es Aix-Marseille I,\\ Aix-Marseille II, et du Sud
Toulon-Var; \\ laboratoire afili\'e \`a la FRUMAM (FR 2291).}

\chapter{The spin-foam-representation of LQG} 
\setcounter{equation}{7}

\begin{abstract}

The problem of background independent quantum gravity is the problem
of defining a quantum field theory of matter and gravity in the
absence of an underlying background geometry.  Loop quantum gravity
(LQG) is a promising proposal for addressing this difficult
task. Despite the steady progress of the field, dynamics remains to a
large extend an open issue in LQG. Here we present the main ideas
behind a series of proposals for addressing the issue of dynamics. We
refer to these constructions as the {\em spin foam representation} of
LQG. This set of ideas can be viewed as a systematic attempt at the
construction of the path integral representation of LQG.

The {\em spin foam representation} is mathematically precise in 2+1
dimensions, so we will start this chapter by showing how it arises in
the canonical quantization of this simple theory. This toy model will
be used to precisely describe the true geometric meaning of the
histories that are summed over in the path integral of generally
covariant theories.

In four dimensions similar structures appear. We call these
constructions {\em spin foam models} as their definition is incomplete
in the sence that at least one of the following issues remains unclear: 1) the
connection to a canonical formulation, and 2) regularization
independence (renormalizability). In the second part of this chapter we will describe the
definition of these models emphasizing the importance of these open issues.
\end{abstract}

\section{The path integral for generally covariant systems}\label{valin}

\begin{figure}[h]\!\!\!\!\!\!
\centerline{\hspace{0.5cm} \(
\begin{array}{c}
\includegraphics[height=5cm]{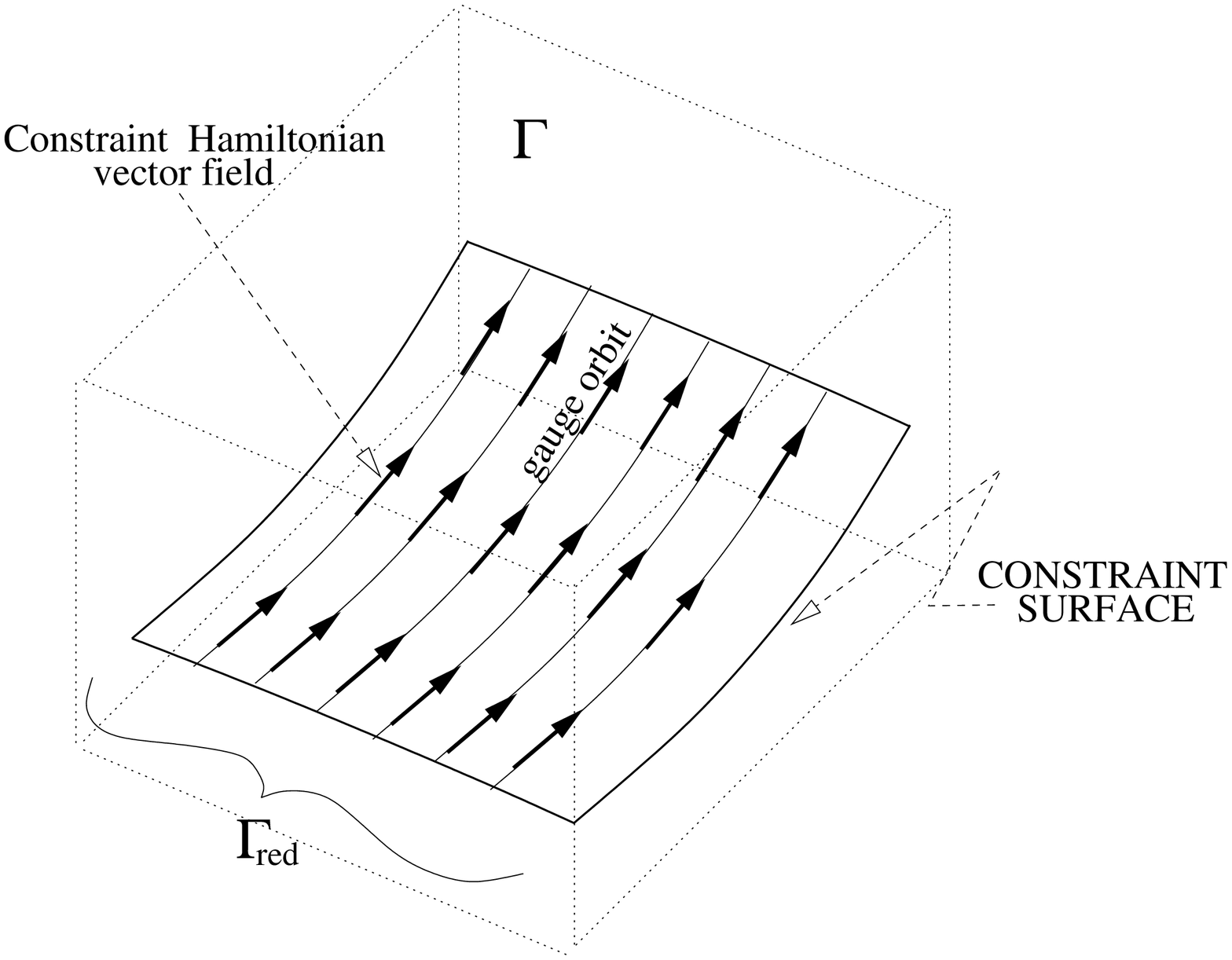}
\end{array}\ \ \ \begin{array}{c}
\includegraphics[height=3cm]{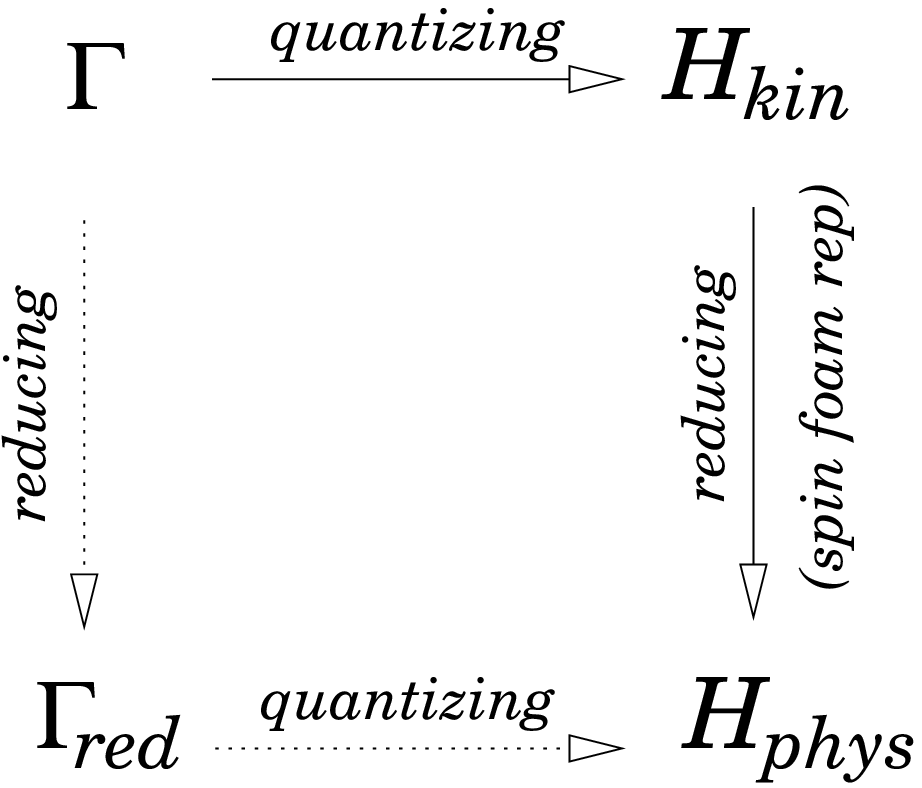}
\end{array}\) } \caption{On the left: the geometry of phase space in gauge
  theories. On the right: the quantization path of LQG (continuous arrows).}
\label{phase}
\end{figure}

LQG is based on the canonical (Hamiltonian) quantization of general relativity
whose gauge symmetry is diffeomorphism invariance. In the Hamiltonian
formulation the presence of gauge symmetries (Dirac P.M.) gives rise to
relationships among the phase space variables---schematically $C(p,q)=0$ for
$(p,q)\in \Gamma$---which are referred to as {\em constraints}. The
constraints restrict the set of possible states of the theory by requiring
them to lay on the constraint hyper-surface. In addition, through the Poisson
bracket, the constraints generate motion associated to gauge transformations on
the constraint surface (see Fig.  (\ref{phase})). The set of physical states
(the so called reduced phase space $\Gamma_{red}$) is isomorphic to the space
of orbits, i.e., two points on the same gauge orbit represent the same state in $\Gamma_{red}$
described in different gauges (Fig.~\ref{phase}). 

In general relativity the absence of a preferred
notion of time implies that the Hamiltonian of gravity
is a linear combination of constraints. This means that Hamilton equations
cannot be interpreted as time evolution and rather correspond to motion along
gauge orbits of general relativity. In generally covariant systems conventional
time evolution is pure gauge: from an initial data satisfying the constraints
one recovers a spacetime by selecting a particular one-parameter family of
gauge-transformations (in the standard ADM context this amounts for choosing a
particular lapse function $N(t)$ and shift $N^a(t)$).

From this perspective the notion of spacetime becomes secondary and
the dynamical interpretation of the the theory seems problematic (in
the quantum theory this is refered to as the {\em ``problem of
time''}).  A possible reason for this apparent problem is the central
role played by the spacetime representation of classical gravity
solutions.  However, the reason for this is to a large part due to the
applicability of the concept of {\em test observers} (or more
generally {\em test fields}) in classical general
relativity\footnote{Most (if not all) of the textbook applications of
general relativity make use of this concept together with the
knowledge of certain exact solutions. In special situations there are
even preferred coordinate systems based on this notion which greatly simplify
interpretation (e.g.  co-moving observers in cosmology, or
observers at infinity for isolated systems).}. Due to the fact that
this idealization is a good approximation to the (classical) process
of observation the notion of spacetime is useful in classical gravity.

As emphasized by Einstein with his hole argument (see Rovelli C. (2005)
for a modern explanation) only the information in relational
statements (independent of any spacetime representation) have physical
meaning. In classical gravity it remains useful to have a spacetime
representation when dealing with idealized test observers. For
instance to solve the geodesic equation and then ask {\em
diff-invariant-questions} such as: {what is the proper time
elapsed on particle 1 between two succesive crossings with particle
2?}  However, already in the classical theory the advantage of the
spacetime picture becomes, by far, less clear if the test particles are
replaced by real objects coupling to the gravitational field
\footnote{In this case one would need first to solve the constraints
of general relativity in order to find the initial data representing
the self-gravitating objects. Then one would have essentially two
choices: 1) Fix a lapse $N(t)$ and a shift $N^a(t)$, evolve with
the constraints, obtain a spacetime (out of the data) in a particular
gauge, and finally ask the {\em diff-invariant-question}; or 2) try to
answer the question by simply studying the data itself (without
t-evolution). It is far from obvious whether the first option (the
conventional one) is any easier than the second.}.

However, this possibility is no longer available in quantum gravity
where at the Planck scale ($\ell_p\approx 10^{-33}cm$) the quantum
fluctuations of the gravitational field become so important that there
is no way (not even in principle\footnote{In order to make a Planck
scale observation we need a Planck energy probe (think of a Planck
energy photon). It would be absurd to suppose that one can disregard
the interaction of such photon with the gravitational field treating
it as test photon.}) to make observations without affecting the
gravitational field. In this context there cannot be any, {\em a
priori}, notion of time and hence no notion of spacetime is possible at
the fundamental level. A spacetime picture would only arise in the
semi-classical regime with the identification of some subsystems that
approximate the notion of test observers.

What is the meaning of the path integral in the background
independent context? The previous discussion rules out the
conventional interpretation of the path integral. There is no meaningful
notion of transition amplitude between states at different times
$t_1>t_0$ or equivalently a notion of {\em ``unitary time
evolution''} represented by an operator $U(t_1-t_0)$.
Nevertheless, a path integral representation of generally
covariant systems arises as a tool for implementing the
constraints in the quantum theory as we argue below.

Due to the difficulty associated with the explicit description of the
reduced phase space $\Gamma_{\rm red}$, in LQG one follows Dirac's
prescription. One starts by quantizing unconstrained phase space
$\Gamma$, representing the canonical variables as self-adjoint
operators in a {\em kinematical Hilbert space} $\Hk$. Poisson brackets
are replaced by commutators in the standard way, and the constraints
are promoted to self-adjoint operators (see Fig.~\ref{phase}). If
there are no anomalies the Poisson algebra of classical constraints is
represented by the commutator algebra of the associated quantum constraints.  In
this way the quantum constraints become the infinitesimal generators
of gauge transformations in $\Hk$. The physical Hilbert space $\Hp$ is
defined as the kernel of the constraints, and hence associated to 
{\em gauge invariant} states. Assuming for simplicity that there is
only one constraint we have \[\psi \in \Hp\ \ \ {\rm iff}\ \ \
\exp[iN \hat C]|\psi\rangle=|\psi\rangle\ \ \
\forall \ \ \ N\in\R,\] where $U(N)=\exp[iN \hat C]$ is the unitary operator
associated to the gauge transformation generated by the constraint $C$ with
parameter $N$. One can characterize the set of gauge invariant states, and
hence construct $\Hp$, by appropriately defining a notion of `averaging' along
the orbits generated by the constraints in $\Hk$. For instance if one can make
sense of the {\em projector} \be P:\Hk\rightarrow \Hp \ \ \ {\rm where } \ \ \
P:=\int dN\ U(N).\label{pipi}\ee It is apparent from the definition that for any $\psi\in
\Hk$ then $P\psi\in\Hp$. The path integral representation arises in the
representation of the unitary operator $U(N)$ as a sum over {\em
gauge-histories} in a way which is technically analogous to standard path
integral in quantum mechanics. The physical interpretation is however quite
different as we will show in Sec.~\ref{gaugy}. The {\em spin foam
representation} arises naturally as the path integral representation of the
field theoretical analog of $P$ in the context of LQG. Needles is to say that
many mathematical subtleties appear when one applies the above formal
construction to concrete examples (Giulini D. \& Marolf D., (1999)).

\section{{\em Spin foams} in 3d quantum gravity}\label{pipo}

Here we derive the {\em spin foam representation} of LQG in a simple
solvable example: 2+1 gravity. For the definition of spin foam
models directly in the covariant picture see Freidel (2005), and
other approaches to 3d quantum gravity see Carlip S. (1998).

\subsection{The classical theory}

Riemannian gravity in $3$ dimensions is a theory with no local
degrees of freedom, i.e., a topological theory. Its action (in the
first order formalism) is given by
\begin{equation}\label{bfaction} S_{}[e,\omega]=\int
\limits_{ M}{\rm Tr}(e\wedge F(\omega)),
\end{equation}
where $M=\Sigma\times \R$ (for $\Sigma$ an arbitrary Riemann
surface), $\omega$ is an $SU(2)$-connection and the triad $e$ is
an $su(2)$-valued $1$-form. The gauge symmetries of the action are
the local $SU(2)$ gauge transformations
\begin{equation}\label{gauge1}
\delta e = \left[e,\alpha \right], \ \ \ \ \ \ \ \ \ \delta \omega
= d_{\omega} \alpha,
\end{equation}
where $\alpha$ is a ${{su(2)}}$-valued $0$-form, and the
`topological' gauge transformation
\begin{equation}\label{gauge2}
\delta e = d_{\va \omega} \eta, \ \ \ \ \ \ \ \ \ \delta \omega =
0,
\end{equation}
where $d_{\va \omega}$ denotes the covariant exterior derivative
and $\eta$ is a ${\tt su(2)}$-valued $0$-form. The first
invariance is manifest from the form of the action, while the
second is a consequence of the Bianchi identity, $d_{\va
\omega}F(\omega)=0$. The gauge symmetries are so large that all
the solutions to the equations of motion are locally pure gauge.
The theory has only global or topological degrees of freedom.

Upon the standard 2+1 decomposition, the phase space in these
variables is parametrized by the pull back to $\Sigma$ of $\omega$
and $e$. In local coordinates one can express them in terms of the
2-dimensional connection $A_a^{i}$ and the triad field
$E^b_j=\epsilon^{bc} e^k_c \delta_{jk}$ where $a=1,2$ are space
coordinate indices and $i,j=1,2,3$ are $su(2)$ indices. The Poisson bracket is given by \be\{A_a^{i}(x),
E^b_j(y)\}=\delta_a^{\, b} \; \delta^{i}_{\, j} \;
\delta^{(2)}(x,y).\end{equation} Local symmetries of the theory
are generated by the first class constraints \be D_b E^b_j = 0, \
\ \ F_{ab}^i(A) = 0, \end{equation} which are referred to as the
Gauss law and the curvature constraint respectively. This
simple theory has been quantized in various ways in the
literature, here we will use it to introduce the {\em
spin foam representation}.

\subsection{{\em Spin foams} from the Hamiltonian
formulation}\label{SFH}

The physical Hilbert space, $\Hp$, is defined by those `states in
$\Hk$' that are annihilated by the constraints. As discussed in
Thiemann (2005) (see also Rovelli C. (2005) and Thiemann (2005)), spin network
states solve the Gauss constraint---$\widehat{ D_a
E^a_i}|s\rangle=0$---as they are manifestly $SU(2)$ gauge invariant. To
complete the quantization one needs to characterize the space of
solutions of the quantum curvature constraints $\widehat
F^i_{ab}$, and to provide it with the physical inner product. As discussed in
Sec.~\ref{valin} we
can achieve this if we can make sense of the following formal
expression for the generalized projection operator $P$:
\be\PP=\int D[N] \ {\rm exp}(i\int
 \limits_{\Sigma} {\rm Tr}[ N \widehat{F}(A)])=\prod \limits_{x\subset \Sigma}
 \delta[\widehat {F(A)}],\label{pis}
\end{equation} where $N(x)\in {\rm su(2)}$. Notice that this is just the field
theoretical analog of equation (\ref{pipi}). $P$ will be defined below
by its action on a dense subset of test-states called the cylindrical
functions ${\rm Cyl}\subset \Hk$ (see Ashtekar \& Lewandowski (2004)). If $P$ exists
then we have \be\label{exists} \langle s
PU[N],s^{\prime}\rangle=\langle s P,s^{\prime}\rangle\ \forall\ \
s,s^{\prime}\in {\rm Cyl}, \ N(x)\in su(2)\ee where $U[N]=\exp(i\int
{\rm Tr}[N\hat F(A)])$. $P$ can be viewed as a map $P:{\rm
Cyl}\rightarrow K_F\subset {\rm Cyl}^{\star}$ (the space of linear
functionals of $\rm Cyl$) where $K_F$ denotes the kernel of the
curvature constraint.  The physical inner product is defined as
\be\label{meo} \langle s^{\prime},s\rangle_p:=\langle s^{\prime} P,s
\rangle,\ee where $\langle,\rangle$ is the inner product in $\Hk$, and
the physical Hilbert space as \be\Hp:={\overline{{\rm Cyl}/J}} \ \ \
{\rm for}\ \ \ J:=\{s \in {\rm Cyl}\ \ {\rm s.t.} \ \ \langle
s,s\rangle_p=0\}\label{null},\ee where the bar denotes the standard
Cauchy completion of the quotient space in the physical norm.

One can make (\ref{pis}) a rigorous definition if one introduces a
regularization.  A regularization is necessary to avoid the naive UV
divergences that appear in QFT when one quantizes non-linear
expressions of the canonical fields such as $F(A)$ in this case (or
those representing interactions in standard particle physics). A
rigorous quantization is achieved if the regulator can be removed
without the appearance of infinities, and if the number of ambiguities
appearing in this process is under control (more about this in
Sec.~\ref{amby}). We shall see that all this can be done in the simple
toy example of this section.

\begin{figure}[h]
\centerline{\hspace{0.5cm} \(
\begin{array}{c}
\includegraphics[width=6cm]{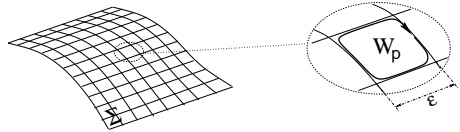}
\end{array}
\) }
\caption{Cellular decomposition of the space manifold $\Sigma$ (a
square lattice of size $\epsilon$ in this example), and the infinitesimal plaquette
holonomy $W_p[A]$.} \label{regu}
\end{figure}

\begin{figure}[h!!!!!]
 \centerline{\hspace{0.0cm} \( \sum \limits_{j k} (2j+1)(2k+1)\!\!\!\!
\begin{array}{c}
\includegraphics[height=2cm]{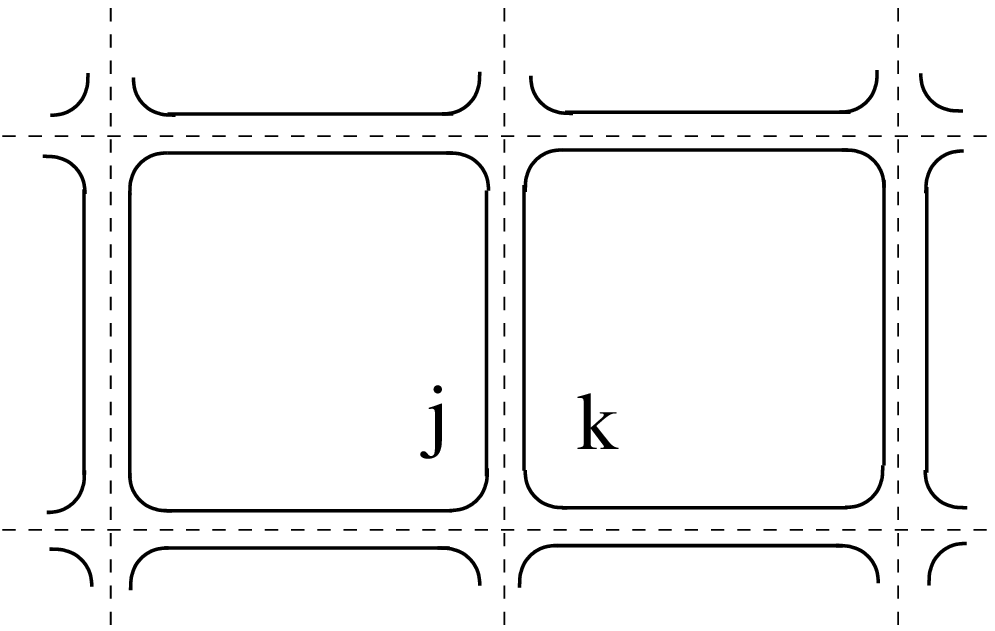}
\end{array} \!\!\!\! =  \sum \limits_k (2k+1)\!\!\!\!
\begin{array}{c}
\includegraphics[height=2cm]{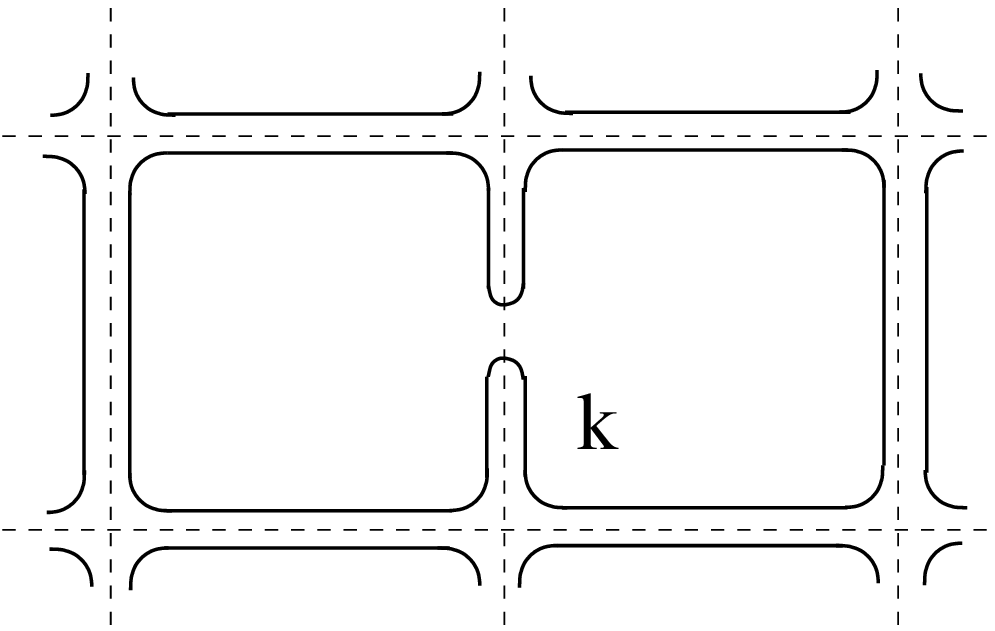}
\end{array}
\) }
\caption{\small In two dimensions the action of two neighboring
 plaquette-sums on the vacuum is equivalent to the action of a single
 larger plaquette action obtained from the fusion of the original
 ones. This implies the trivial {\em scaling} of the physical inner
 product under refinement of the regulator and the existence of a well
 defined limit $\epsilon\rightarrow 0$.}
\label{figurin}
\end{figure}

We now introduce the regularization. Given a partition of $\Sigma$ in terms of
2-dimensional plaquettes of coordinate area $\epsilon^2$ (Fig.~\ref{regu}) one can write the integral
\be \label{***} F[N]:=\int\limits_{\Sigma} {\rm Tr}[ N {F}(A)]=
\lim_{\epsilon\rightarrow 0}\ \sum_{p} \epsilon^2 {\rm
Tr}[N_{p} F_{p}]\end{equation} as a limit of a Riemann sum, where
$N_{p}$ and $F_{p}$ are values of the smearing field $N$ and the curvature
$\epsilon^{ab}F_{ab}^i[A]$ at some interior point of the plaquette
$p$ and $\epsilon^{ab}$ is the Levi-Civita tensor. Similarly the
holonomy $W_{p}[A]$ around the boundary of the plaquette $p$ (see
Figure \ref{regu}) is given by \be W_{p}[A]=\mathbbm{1}+ \epsilon^2
F_{p}(A)+{\cal O}(\epsilon^2).\end{equation} The previous two equations
imply that $F[N]=\lim_{\epsilon\rightarrow 0}\sum_{p} {\rm Tr}[N_pW_p]$, and lead to the following definition: given $s, s^{\prime} \in {\rm Cyl}$ (think of
{\em spin network} states) the physical inner product
(\ref{meo}) is given by \be\label{new} \langle s^{\prime}P,s\rangle :=
\lim_{\epsilon\rightarrow 0} \ \ \langle s \prod_{p} \ \int \ dN_{p}
\ {\rm exp}(i {\rm Tr}[ N_{p} {W}_{p}]), s\rangle.
\end{equation}
The partition is chosen so that the links of
the underlying spin network graphs border the plaquettes. One can
easily perform the integration over the $N_{p}$ using the
identity (Peter-Weyl theorem) \be\label{pw}\int \ dN \
 {\rm exp}(i {\rm Tr}[ N {W}])=\sum_{j} \ (2j + 1) \ {\rm Tr}[\stackrel{j}{\Pi}\!(W)],\end{equation}
where $\stackrel{j}{\Pi}\!(W)$ is the spin $j$ unitary irreducible
representation of $SU(2)$. Using the previous equation \be\label{final}
 \langle s^{\prime}P,s\rangle := \lim_{\epsilon\rightarrow 0} \ \ \prod^{n_p(\epsilon)}_{p} \sum_{j_p}
 (2j_p+1)\ \langle s^{\prime}\
{\rm Tr}[\stackrel{ j_p}{\Pi}\!({W}_{p})]), s\rangle,
\end{equation} where the spin $j_p$ is associated to
 the p-th plaquette, and $n_p(\epsilon)$ is the number of
 plaquettes. Since the elements of the set of Wilson loop operators
 $\{W_p\}$ commute, the ordering of plaquette-operators in the
 previous product does not matter. The limit $\epsilon
\rightarrow 0$ exists and one can give a closed expression for the
physical inner product.  That the regulator can be removed follows
from the orthonormality of $SU(2)$ irreducible representations
which implies that the two spin sums associated to the action of two
neighboring plaquettes collapses into a single sum over the action
of the {\em fusion} of the corresponding plaquettes (see
Fig~\ref{figurin}). One can also show that it is
finite\footnote{The physical inner product between spin network
states satisfies the following inequality \[ \left|
\langle s,s^{\prime}\rangle_p\right|\le C \sum_{j} (2j+1)^{2-2g},\] for some
positive constant $C$. The convergence of the sum for genus $g\ge
2$ follows directly. The case of the sphere $g=0$ and the torus
$g=1$ can be treated individually (Noui K. \& Perez A. (2005)).}, and satisfies all
the properties of an inner product (Noui K. \& Perez A. (2005)).

\subsection{The spin foam representation}

\begin{figure}[h!!!!!!!!!!!!!!!!!!!!!!!!!!!!!!!!!!1]
\centerline{\hspace{0.5cm} \( {\rm Tr}[\stackrel{k}{\Pi}\!(W_{p})]
\rhd \!\!\!\!\!\!\!\!\!\!\!\!\!\!\!\!\begin{array}{c}
\includegraphics[width=3.6cm]{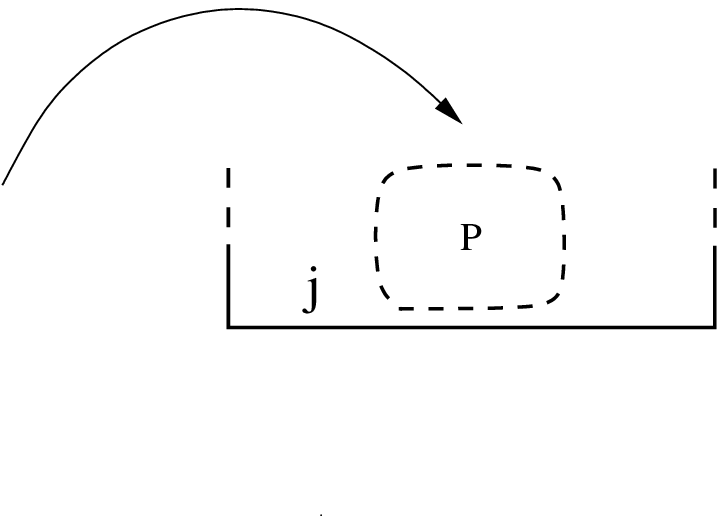}
\end{array}
=
\sum \limits_{m} N_{j,m,k}
\begin{array}{c}\includegraphics[width=2.5cm]{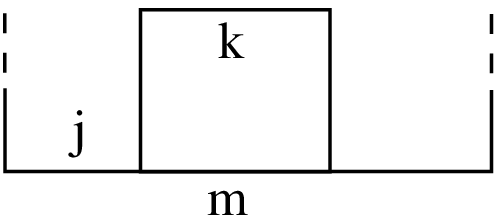}
\end{array}
\) }
\caption{Graphical notation representing the action of one
plaquette holonomy on a {\em spin network} state. On the right is
the result written in terms of the {\em spin network} basis. The
amplitude $N_{j,m,k}$ can be expressed in terms of Clebsch-Gordan
coefficients.} \label{pito}
\end{figure}

\begin{figure}[h!!!!!]
 \centerline{\hspace{0.0cm}\(
\begin{array}{ccc}
\includegraphics[height=1cm]{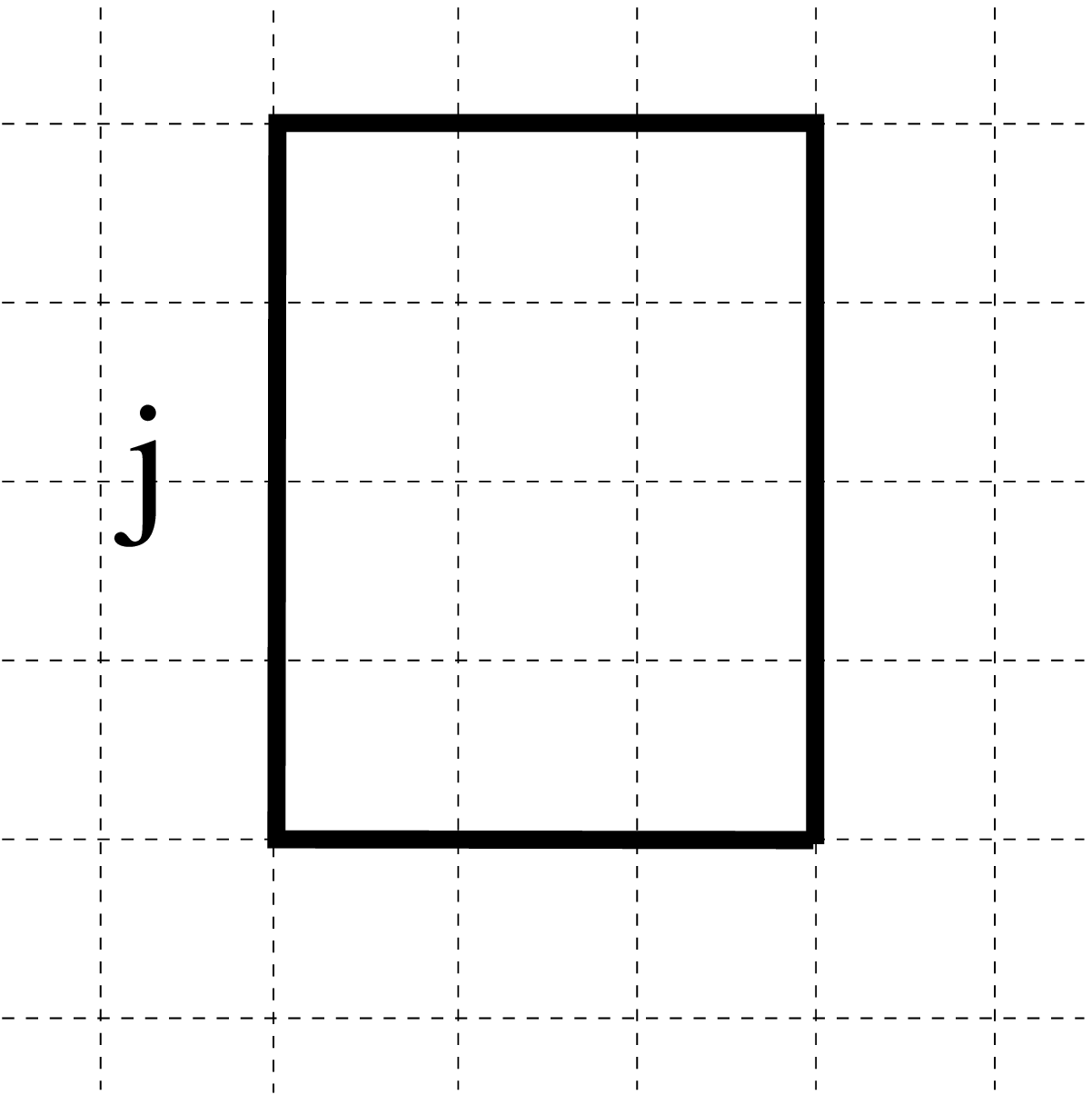} \\
\includegraphics[height=1cm]{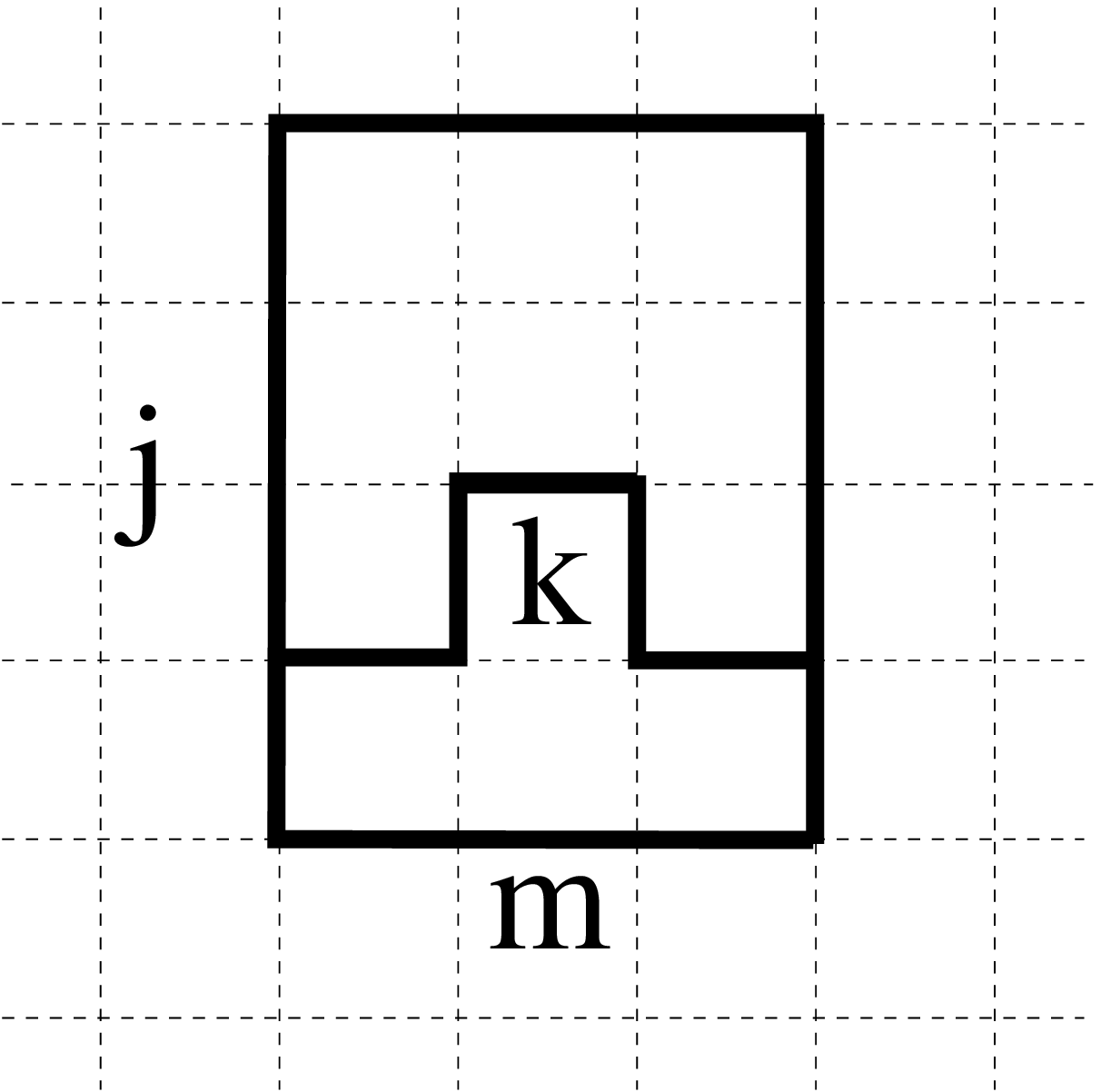}\\
\includegraphics[height=1cm]{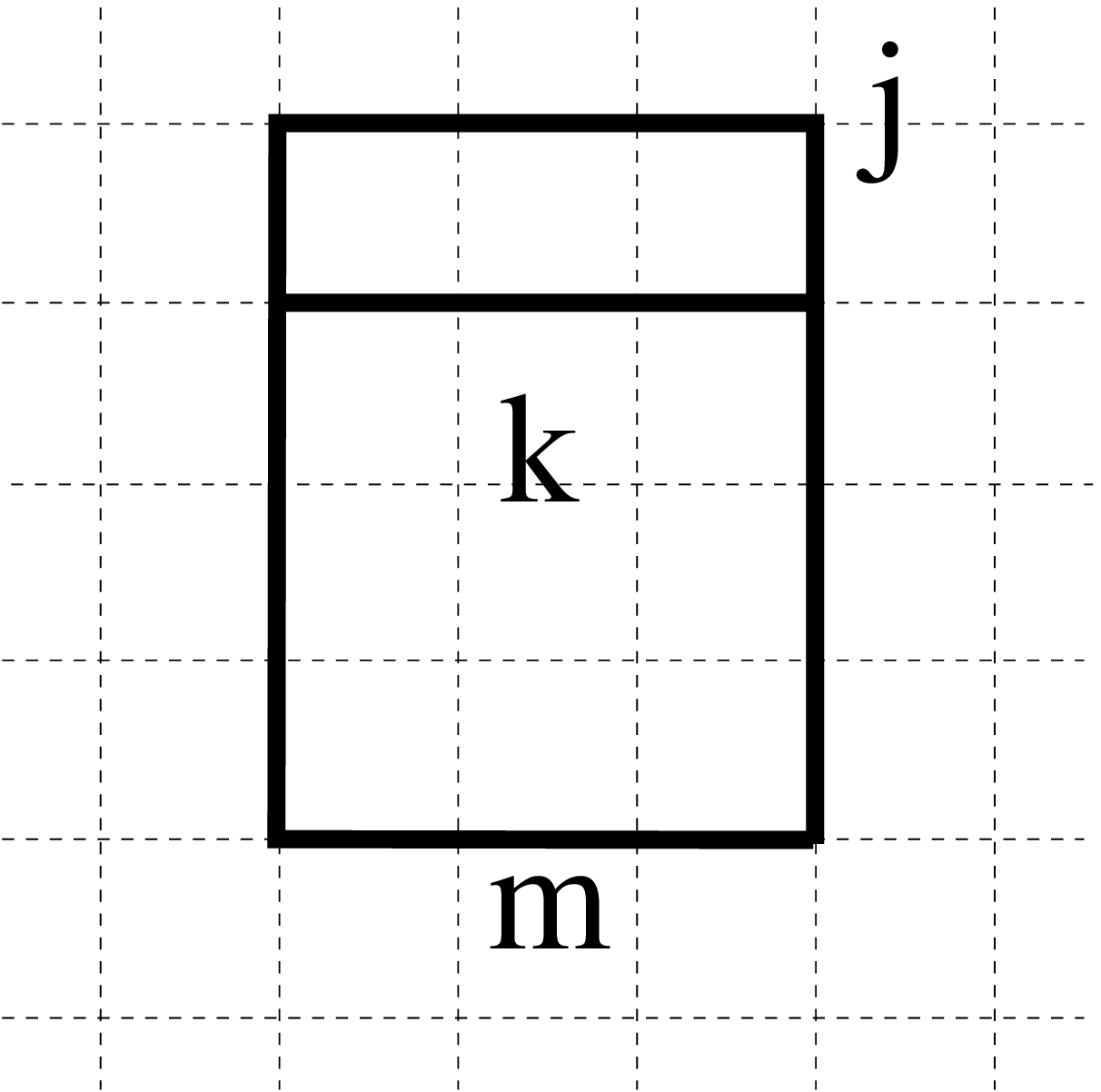}
\end{array}
\begin{array}{ccc}
\includegraphics[height=1cm]{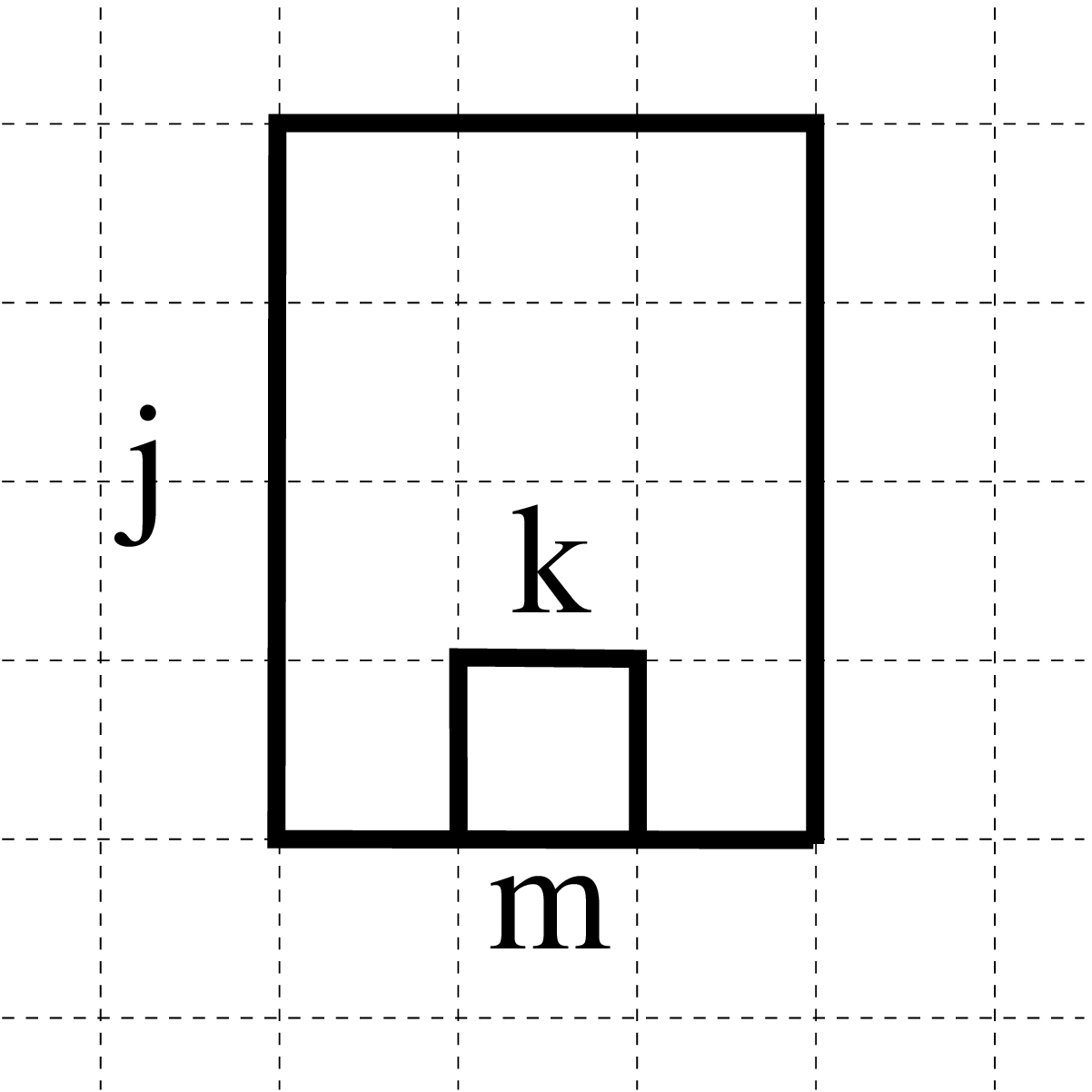} \\
\includegraphics[height=1cm]{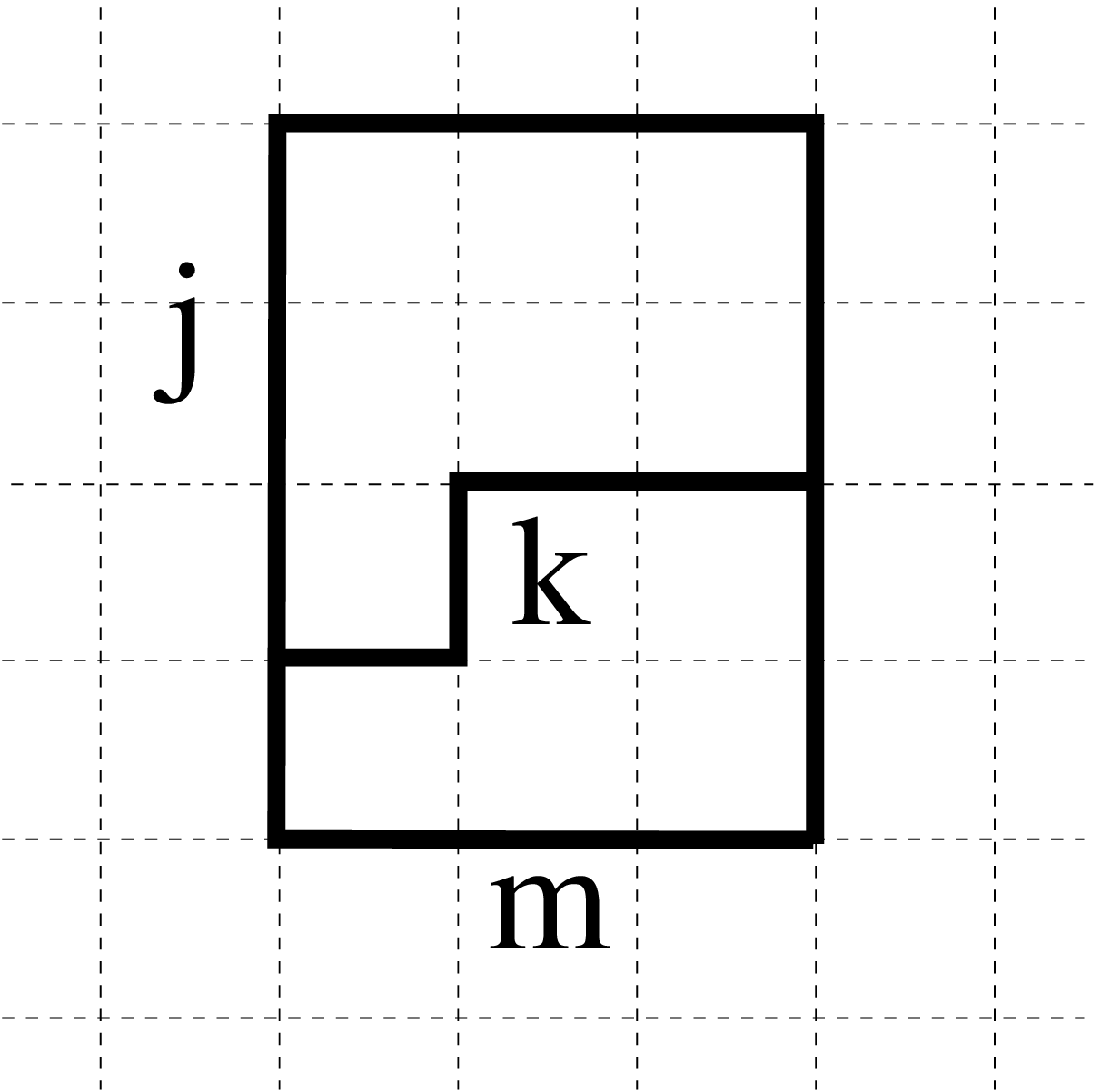}\\
\includegraphics[height=1cm]{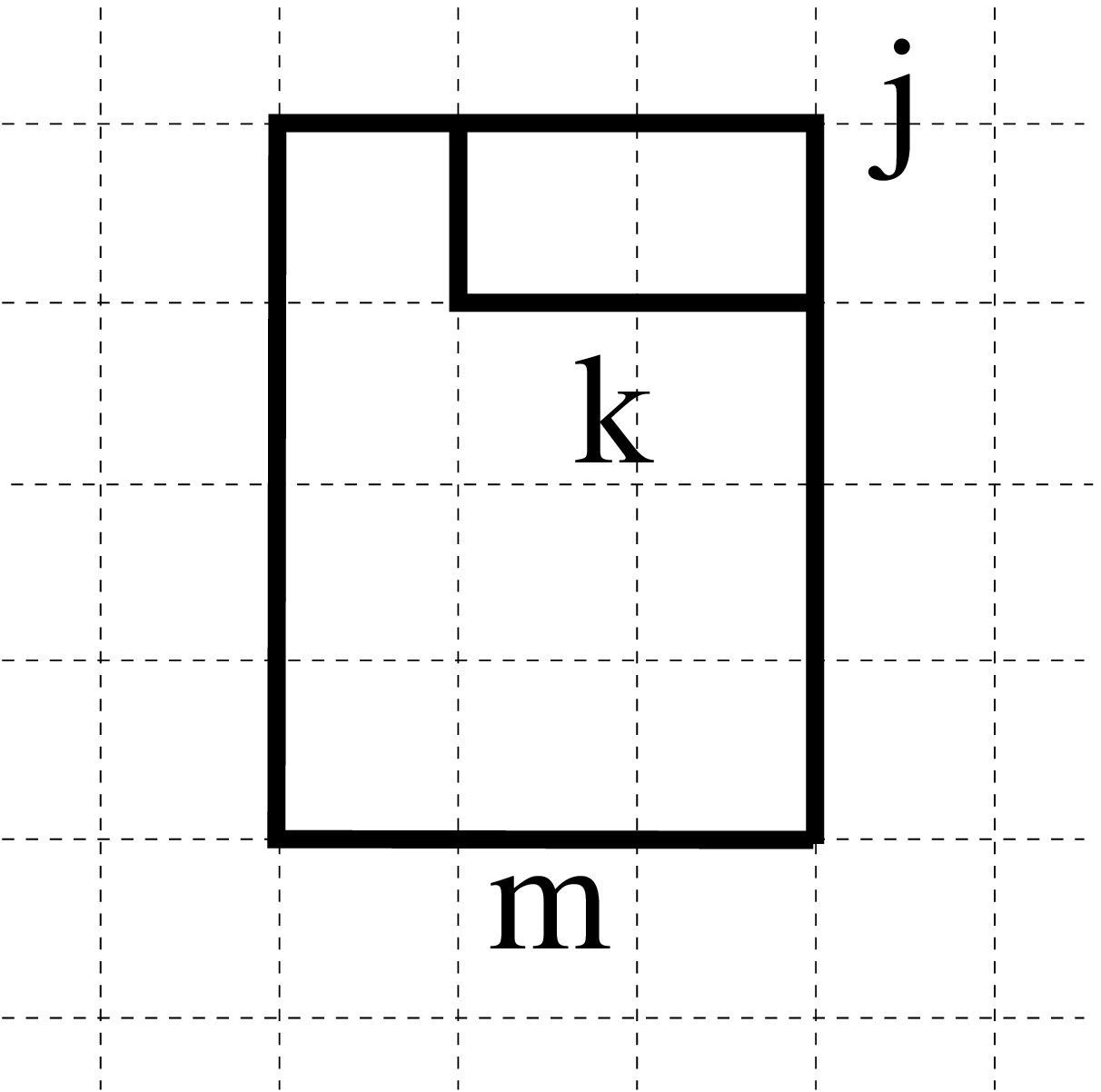}
\end{array}
\begin{array}{ccc}
\includegraphics[height=1cm]{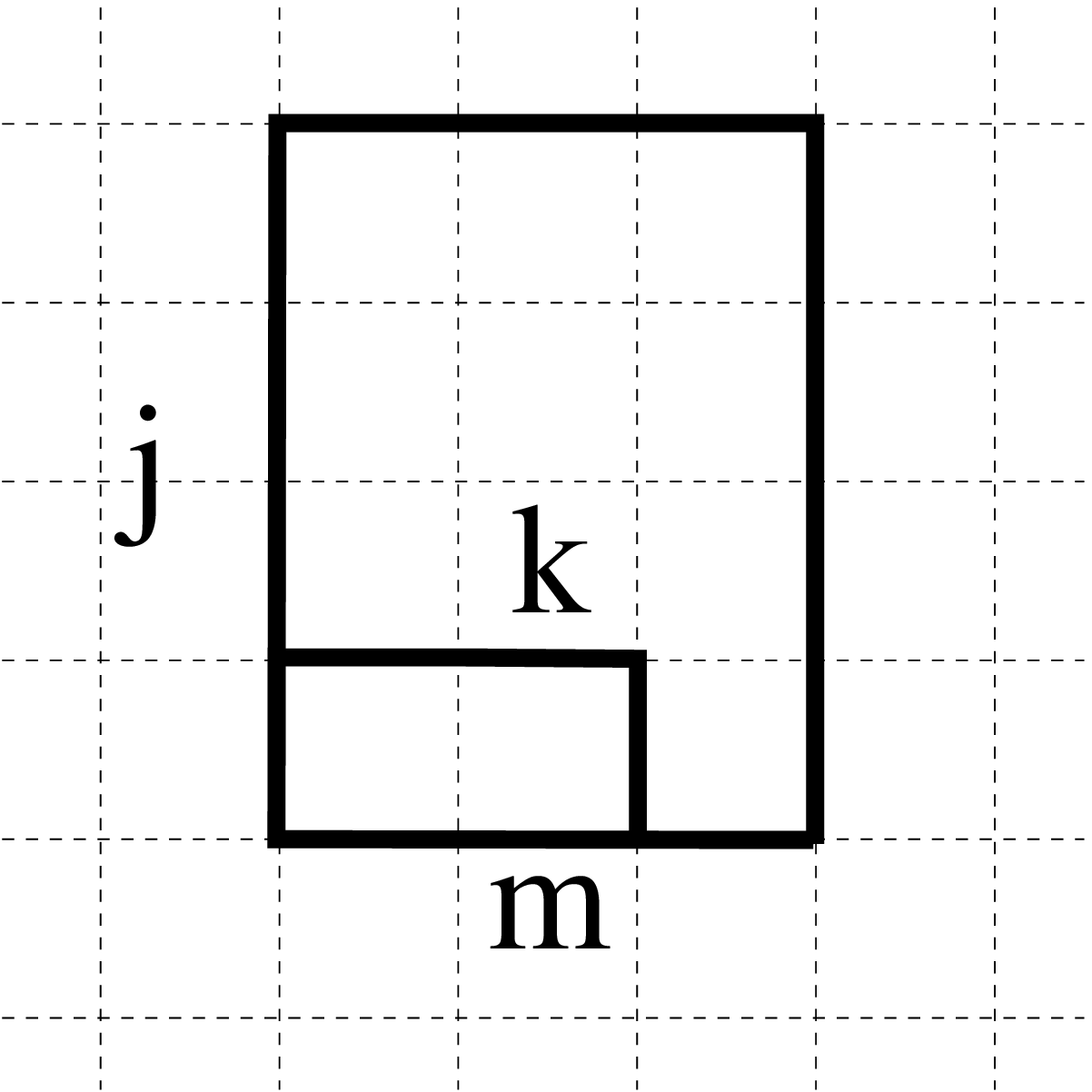} \\
\includegraphics[height=1cm]{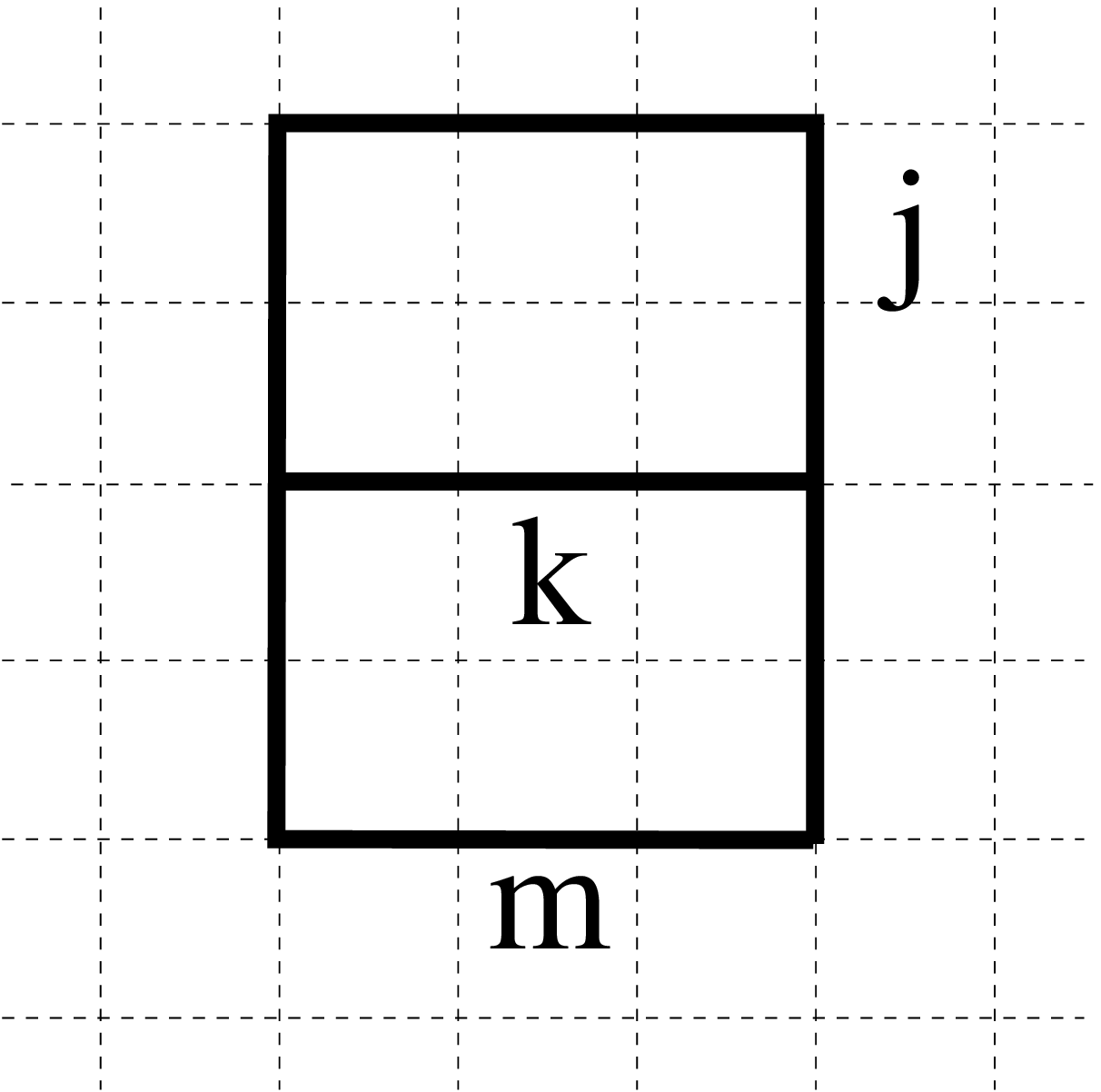}\\
\includegraphics[height=1cm]{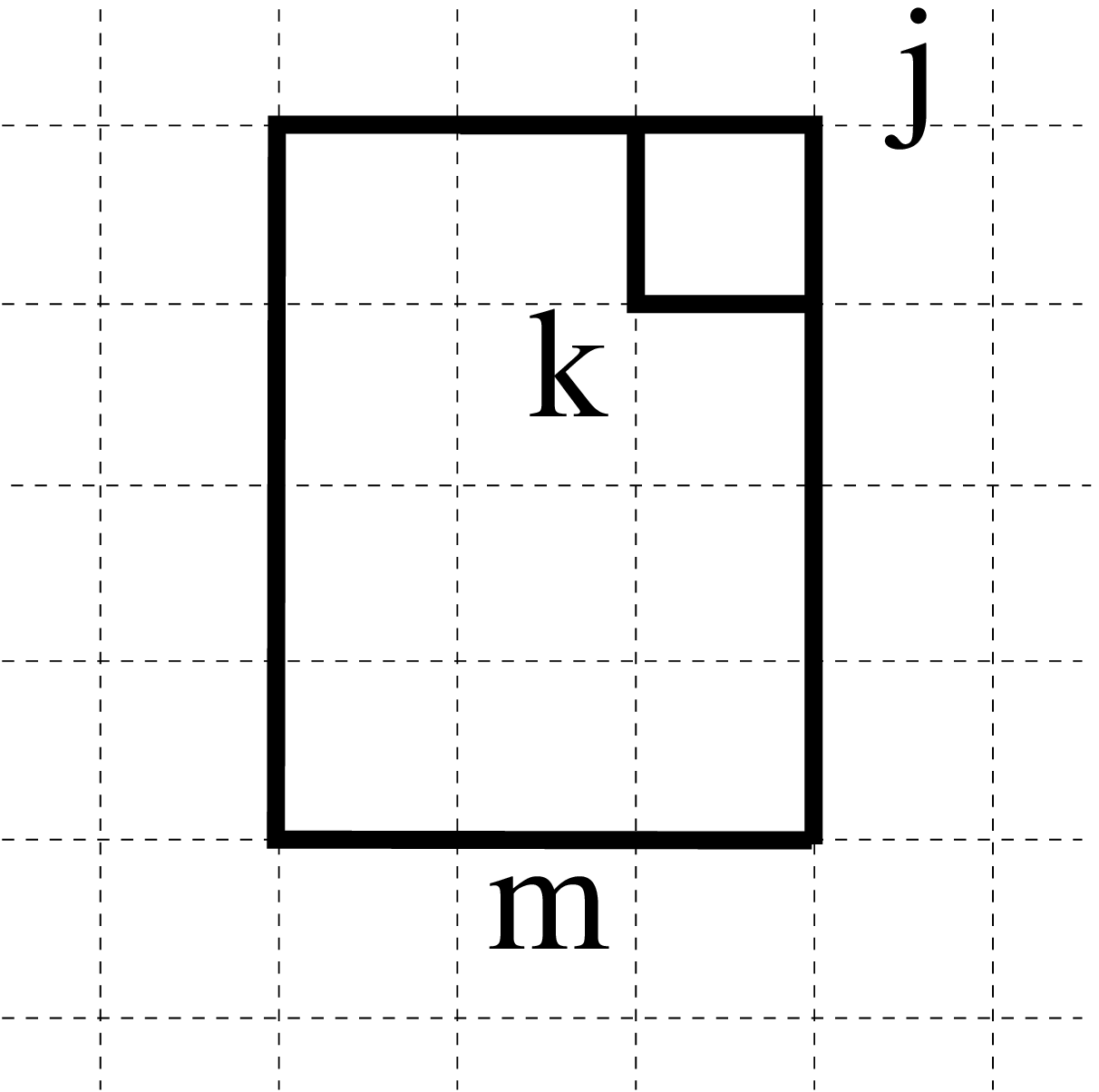}
\end{array}
\begin{array}{ccc}
\includegraphics[height=1cm]{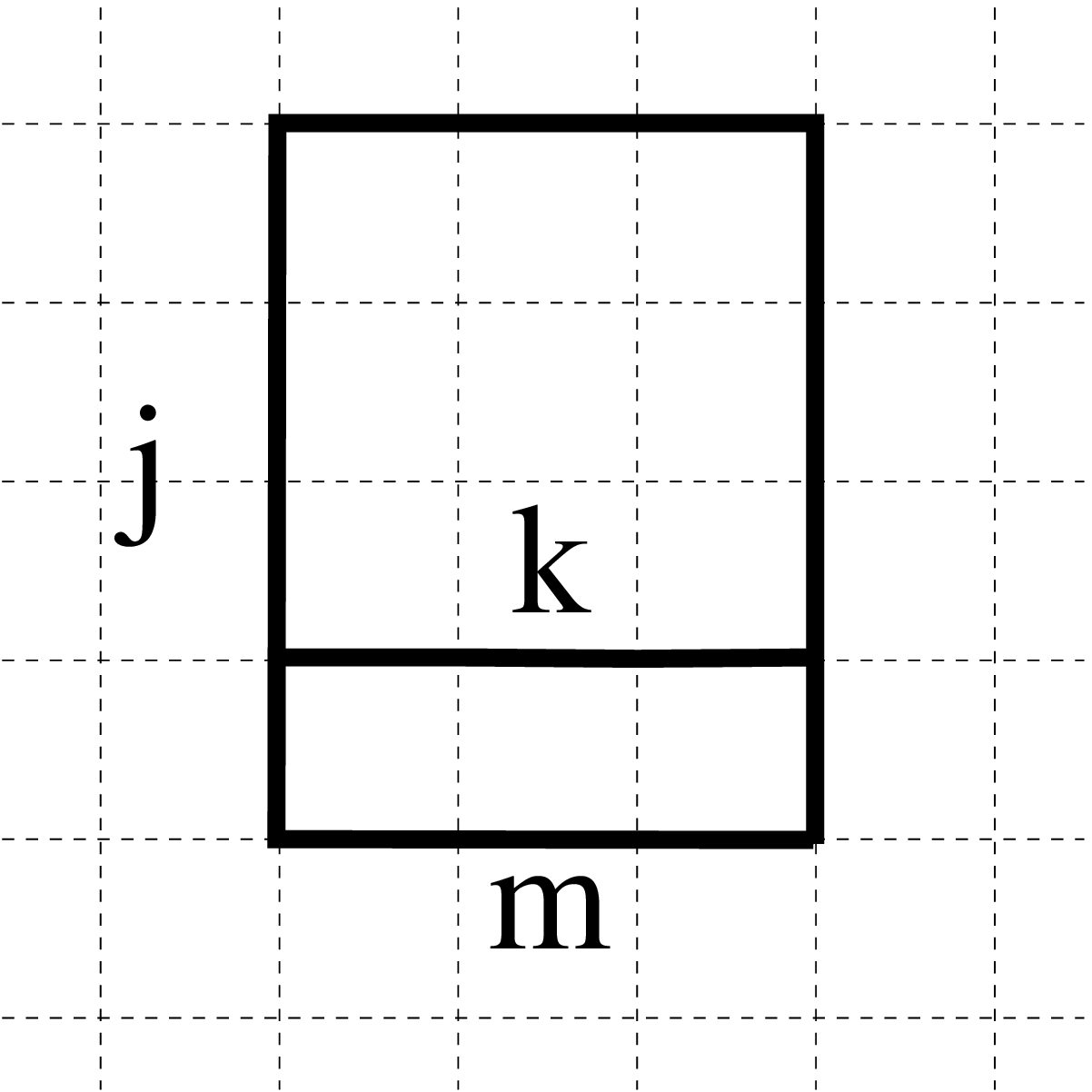} \\
\includegraphics[height=1cm]{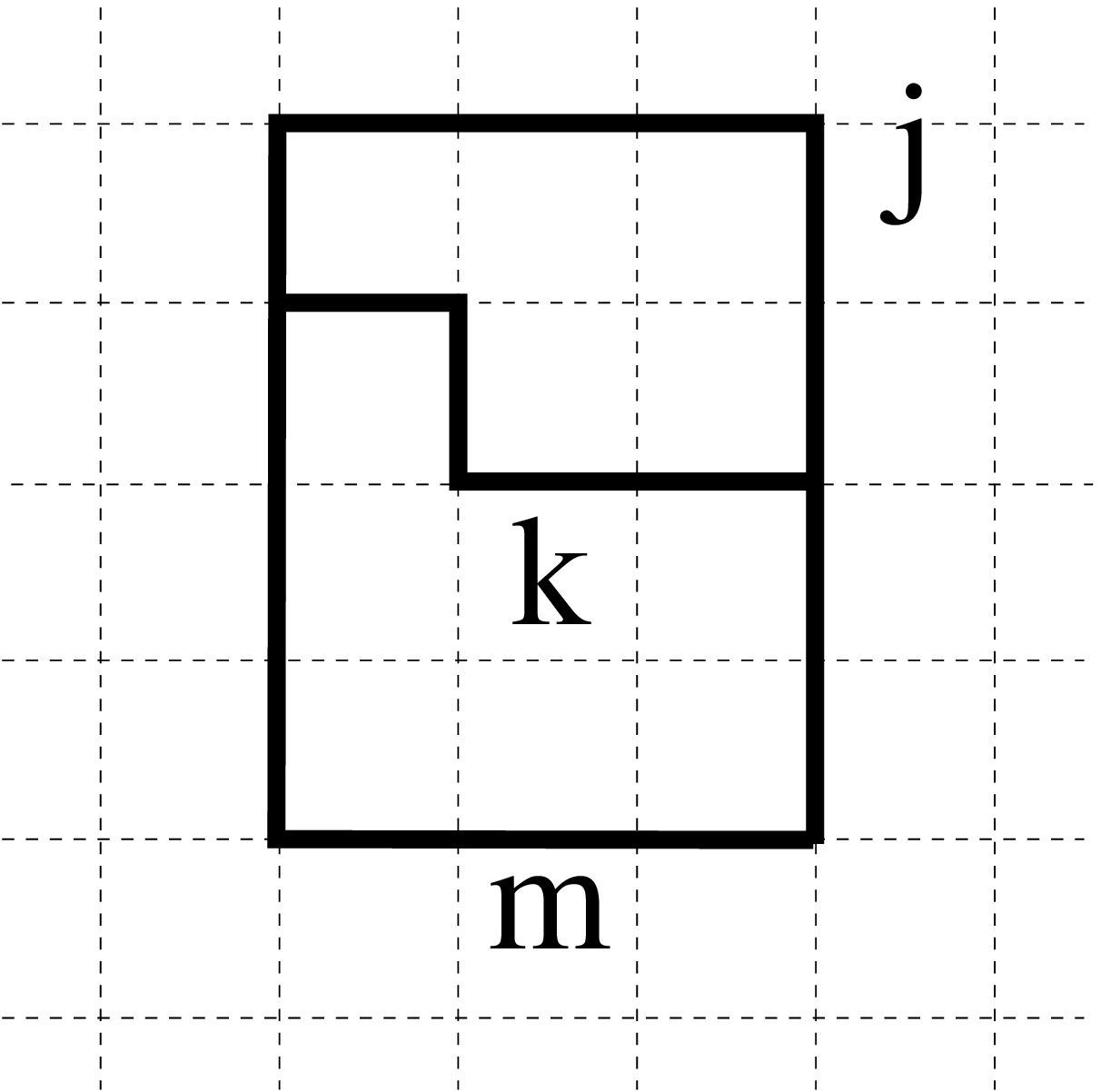}\\
\includegraphics[height=1cm]{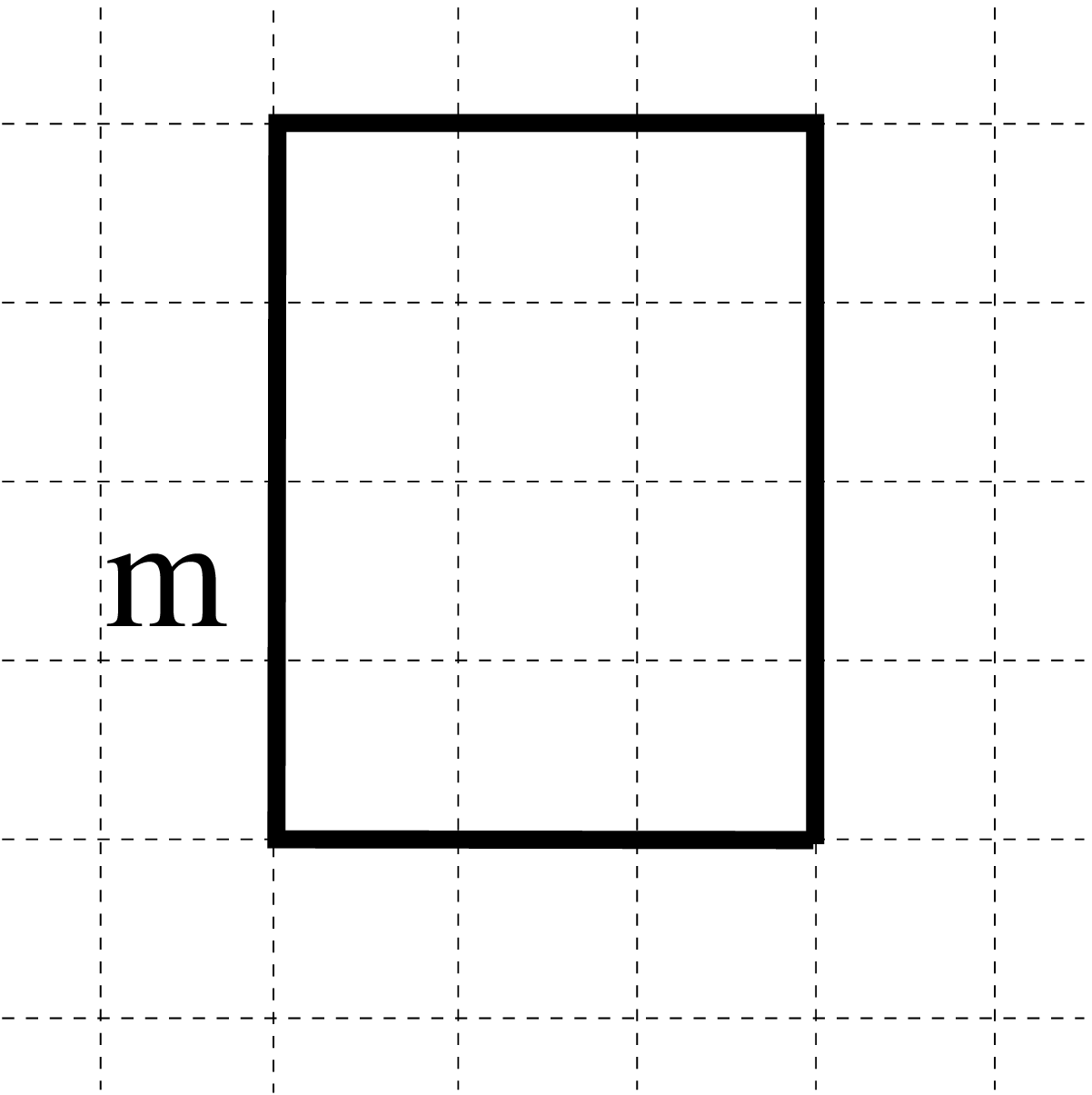}
\end{array}\begin{array}{c}
\includegraphics[height=.3cm]{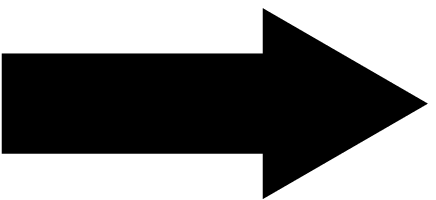}
\end{array}\!\!\!
\begin{array}{c}
\includegraphics[height=3cm]{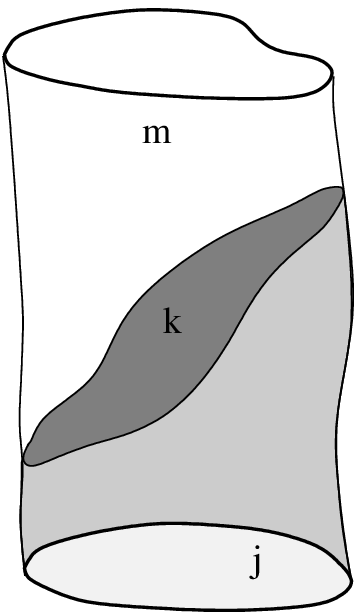}
\end{array}
\)} \caption{\small A set of discrete transitions in the
loop-to-loop physical inner product obtained by a series of
transitions as in Figure \ref{pito}. On the right, the continuous
{\em spin foam representation} in the limit $\epsilon\rightarrow
0$.} \label{lupy}
\end{figure}

\begin{figure}
\centerline{\hspace{0.5cm} \( {\rm
Tr}[\stackrel{n}{\Pi}\!(W_{p})]\rhd
\!\!\!\!\!\!\!\!\!\!\!\!\!\!\!\!\begin{array}{c}
\includegraphics[width=2.5cm]{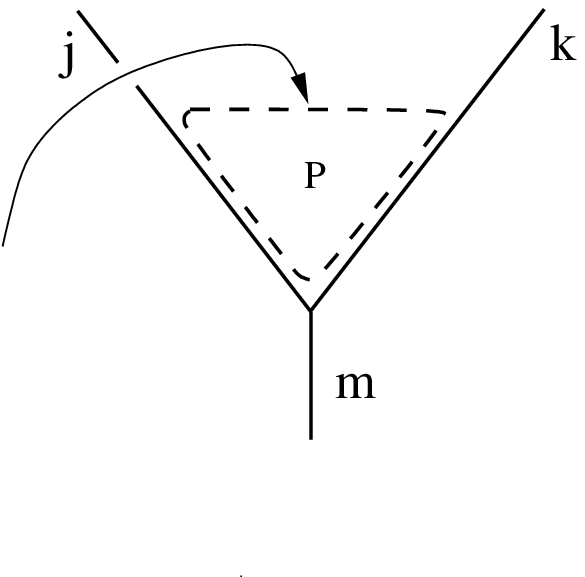}
\end{array}
=
\sum\limits_{o,p} \frac{1}{\Delta_n \Delta_j \Delta_k \Delta_m}
\left\{\begin{array}{ccc}j\ \ k \ \ m\\ n\ \  o\ \  p
\end{array}\right\}
\begin{array}{c}
\includegraphics[width=2.5cm]{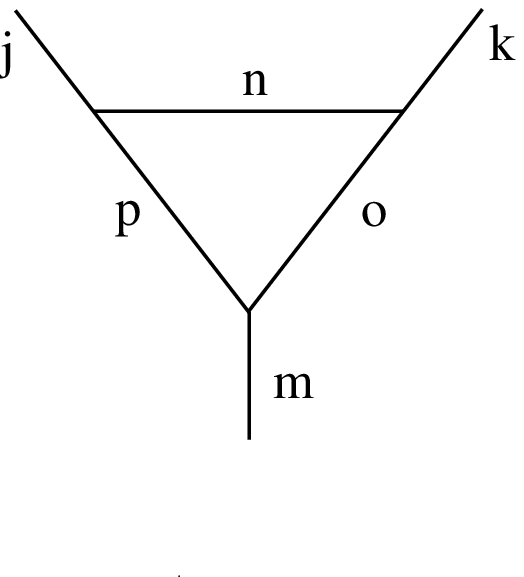}
\end{array}
\) }
\caption{Graphical notation representing the action of one
plaquette holonomy on a {\em spin network} vertex. The object in
brackets ($\{\}$) is a $6j$-symbol and $\Delta_j:=2j+1$.}
\label{pitolon}
\end{figure}

\begin{figure}[h!!!!!!!!!!!!!!!!!!!!!!!!!!!!!!!!!!!!!!!!!!!!!!!!!!!!!!!!!!!!!!!!!!!]
 \centerline{\hspace{0.5cm}\(
\begin{array}{ccc}
\includegraphics[height=1.4cm]{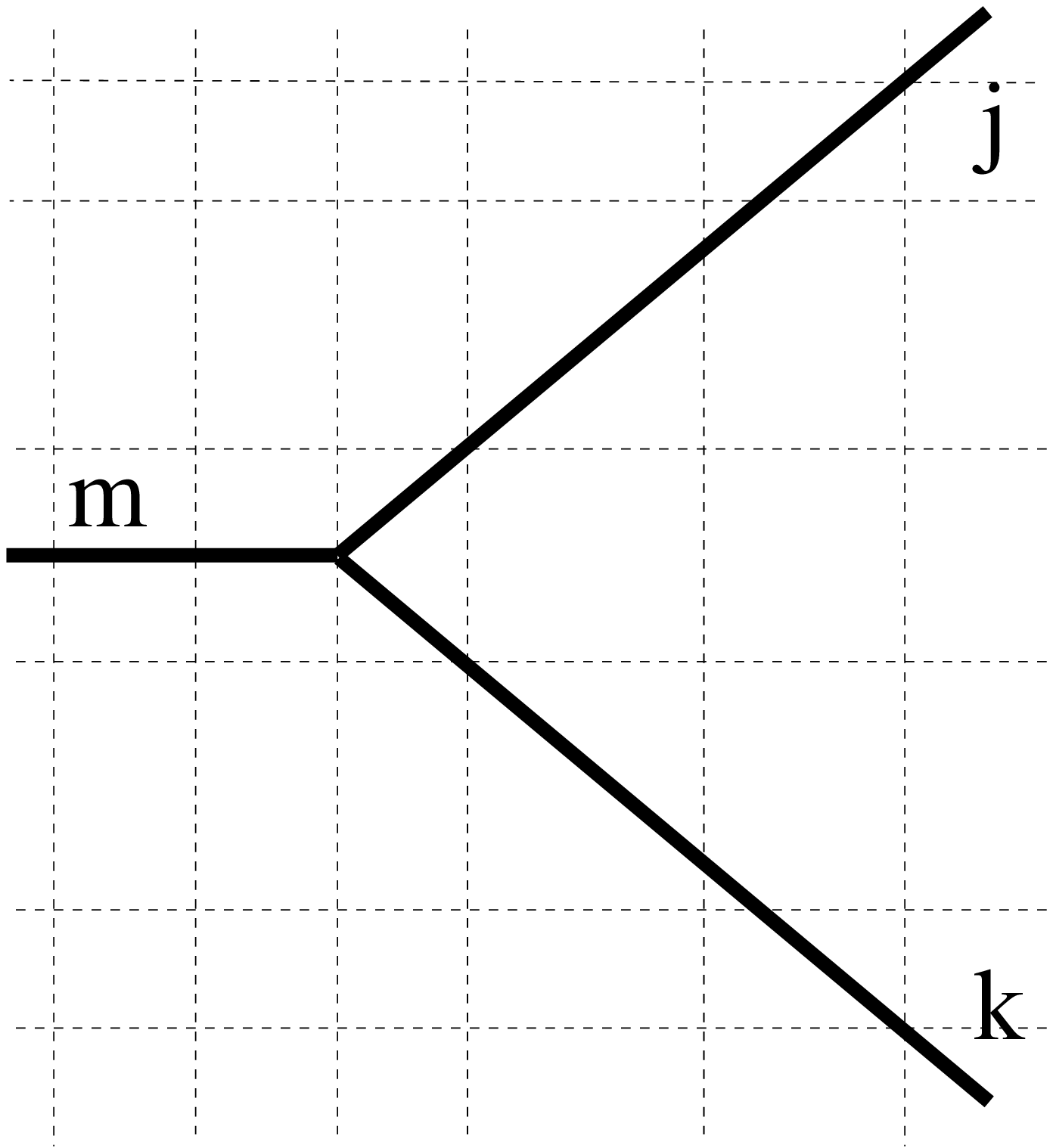} \\
\includegraphics[height=1.4cm]{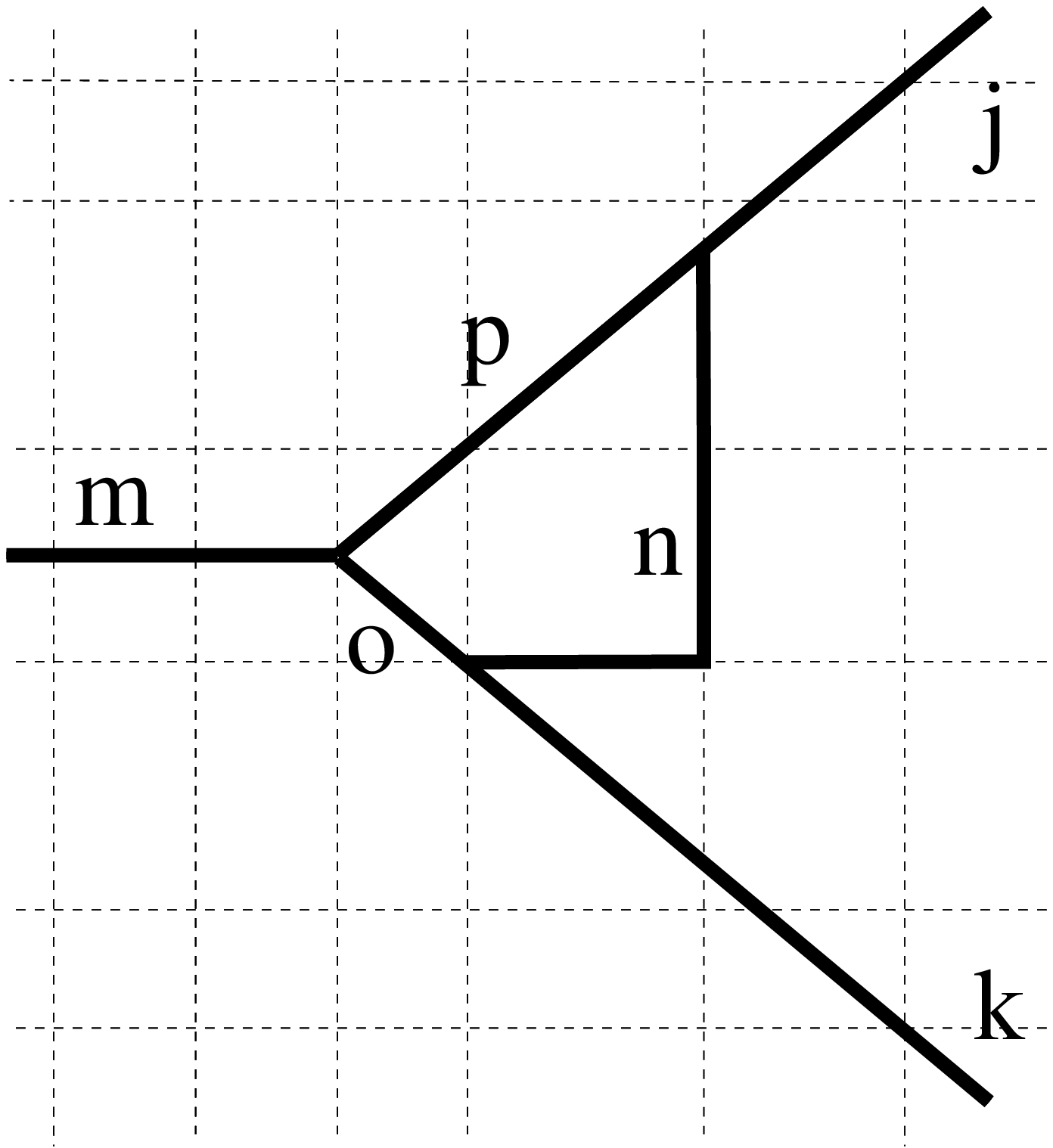} \\
\includegraphics[height=1.4cm]{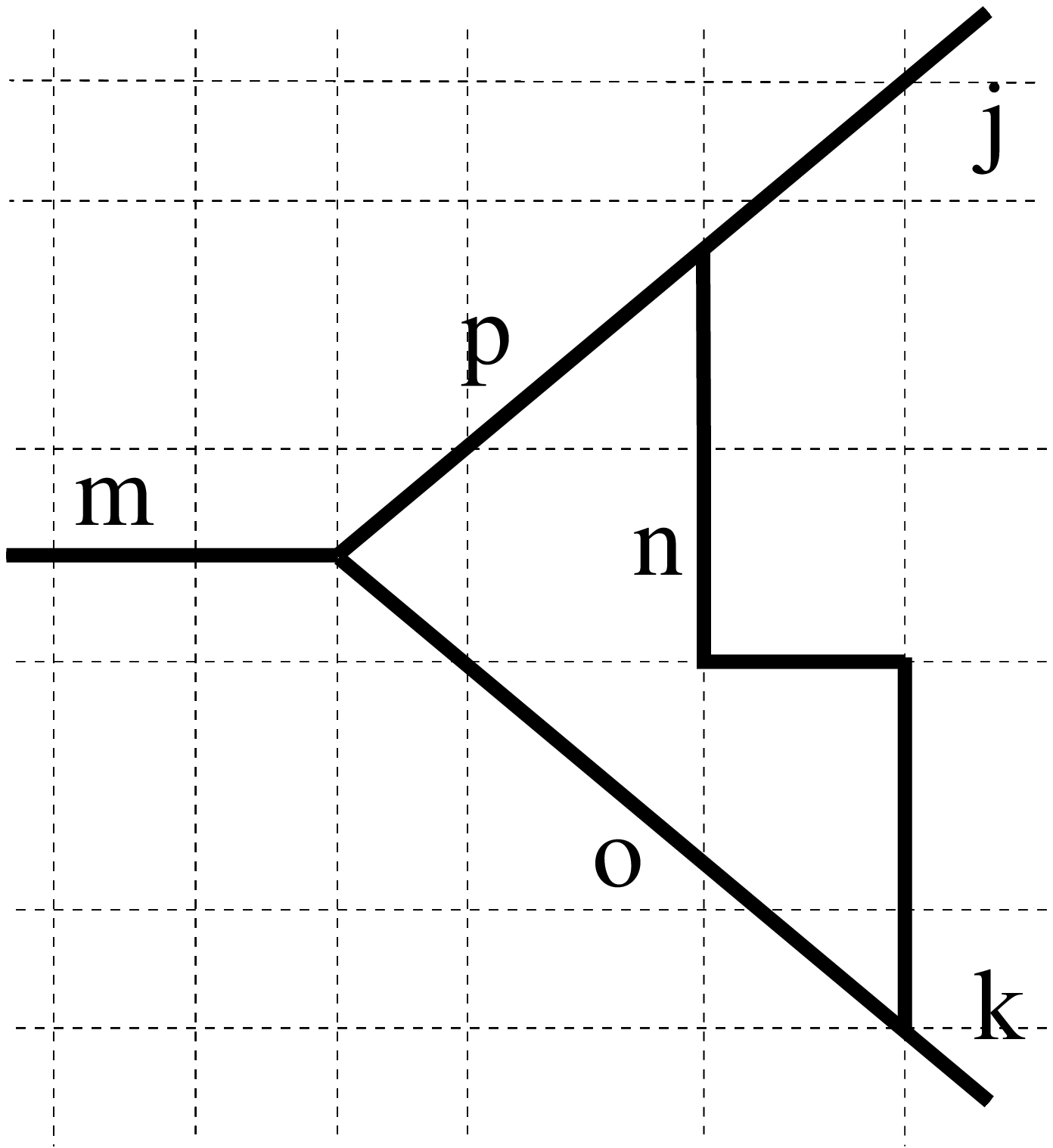}
\end{array}
\begin{array}{c}
\includegraphics[height=1.4cm]{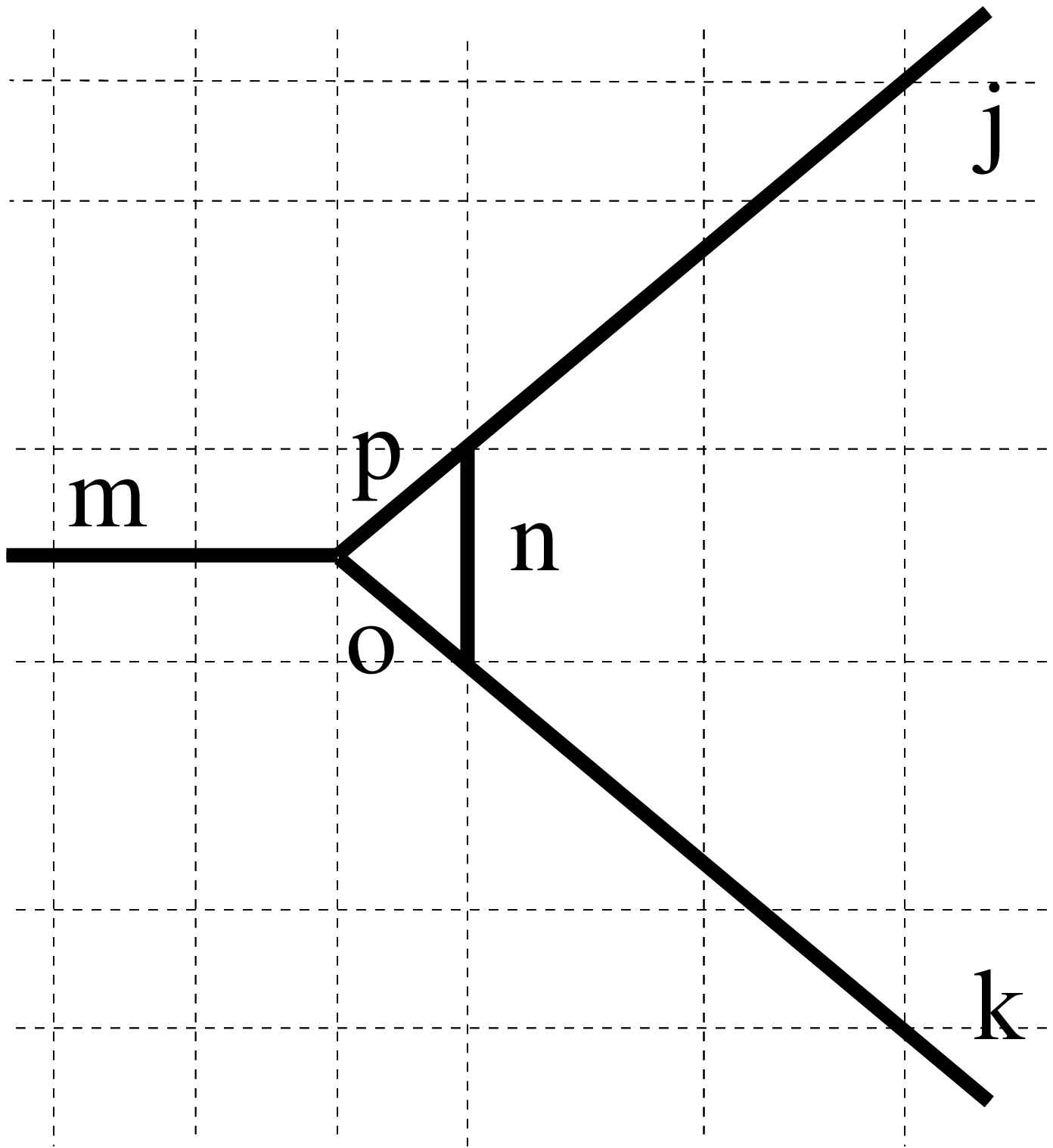}\\
\includegraphics[height=1.4cm]{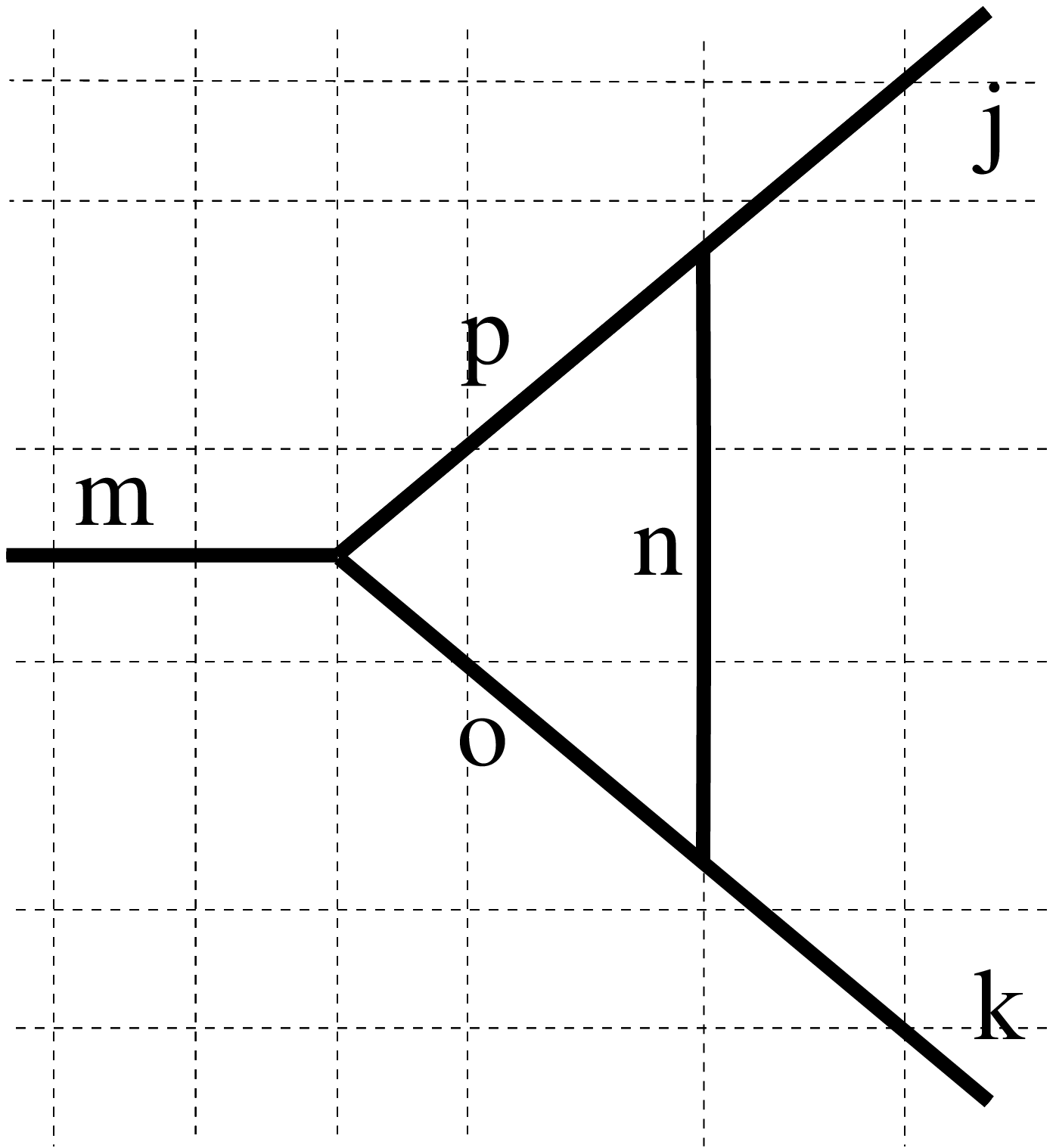}\\
\includegraphics[height=1.4cm]{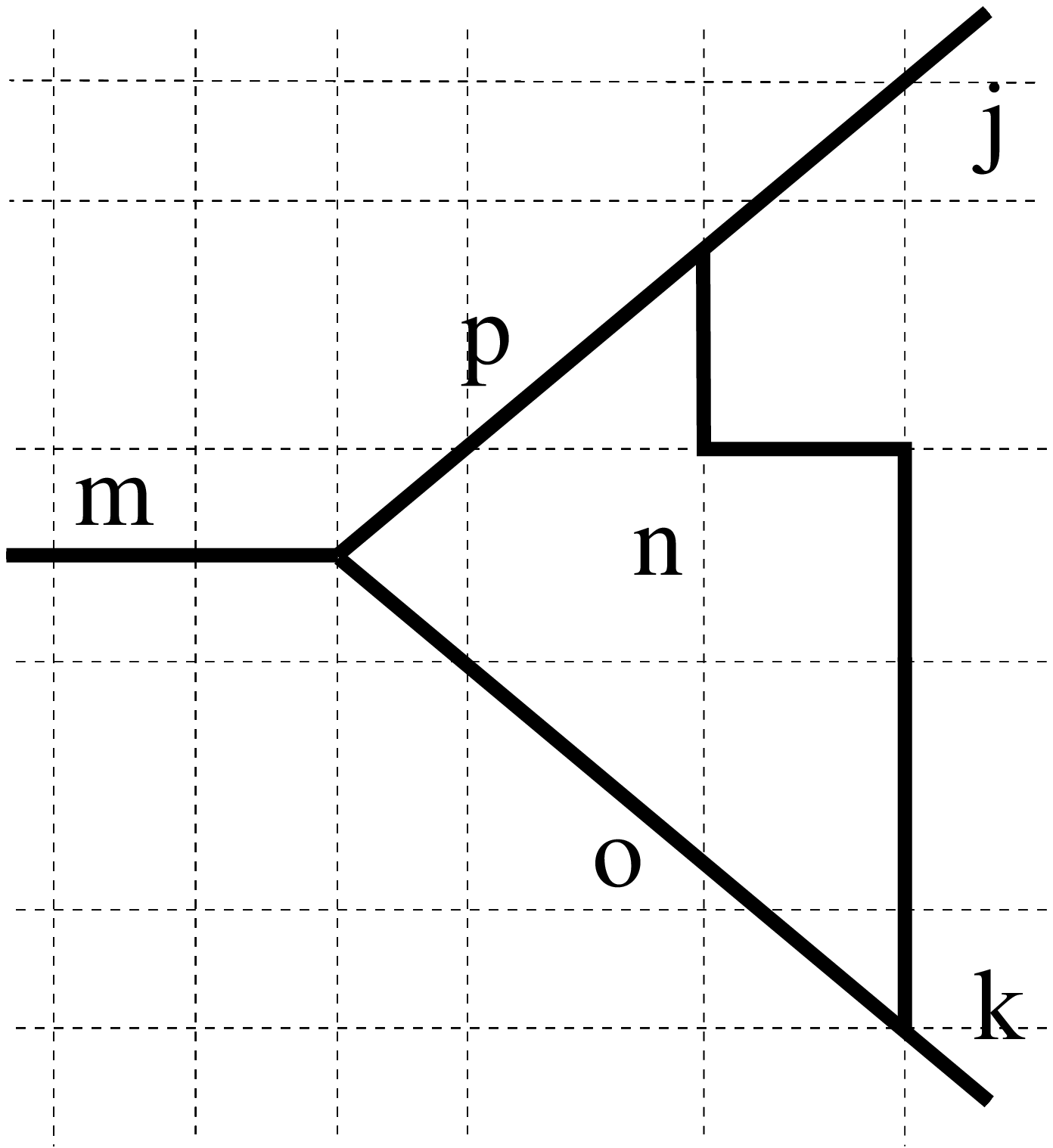}
\end{array}
\begin{array}{c}
\includegraphics[height=1.4cm]{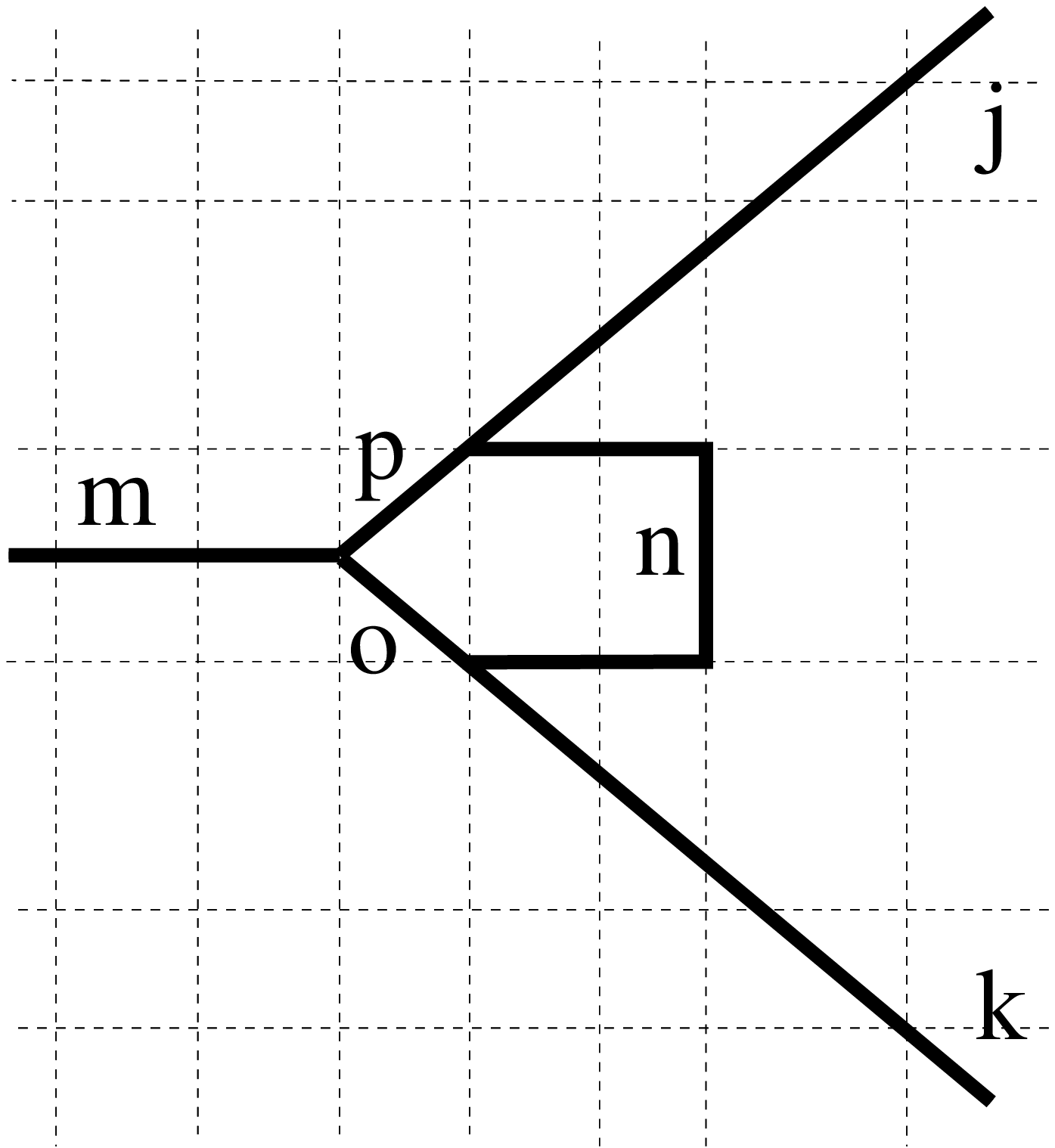}\\
\includegraphics[height=1.4cm]{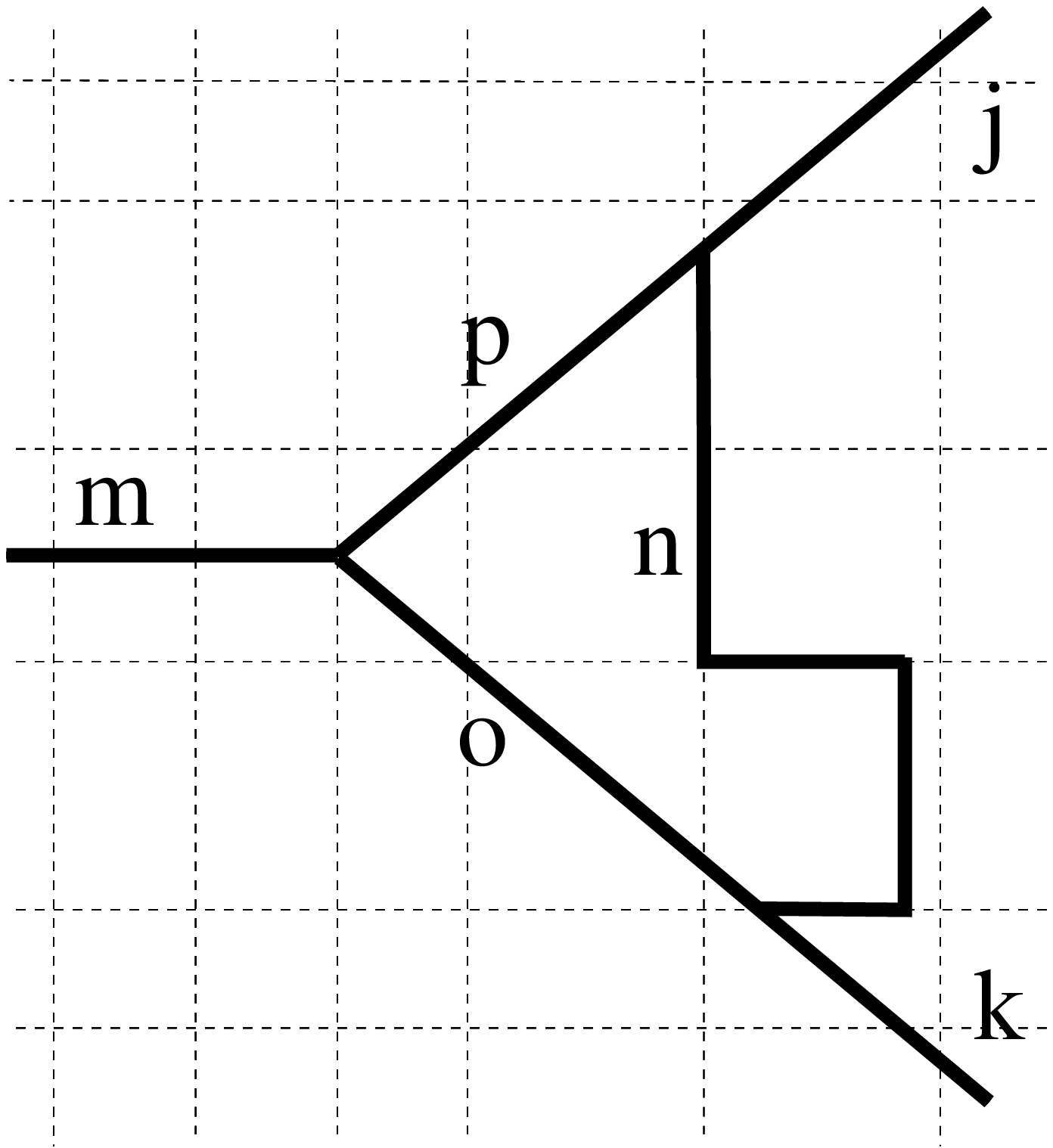}\\
\includegraphics[height=1.4cm]{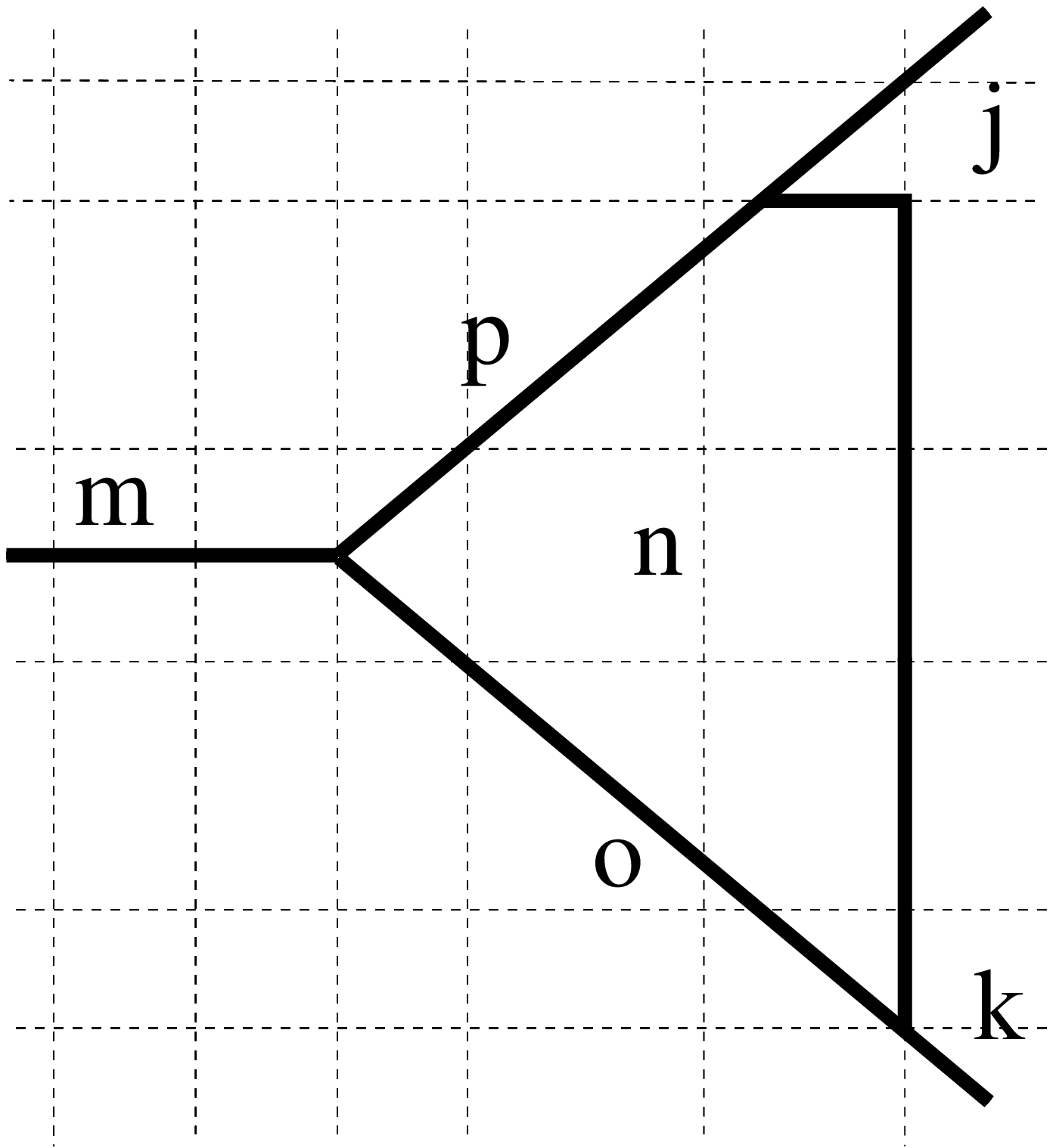}
\end{array}\begin{array}{c}
\includegraphics[height=.3cm]{flecha.eps}
\end{array}
\begin{array}{c}
\includegraphics[height=4cm]{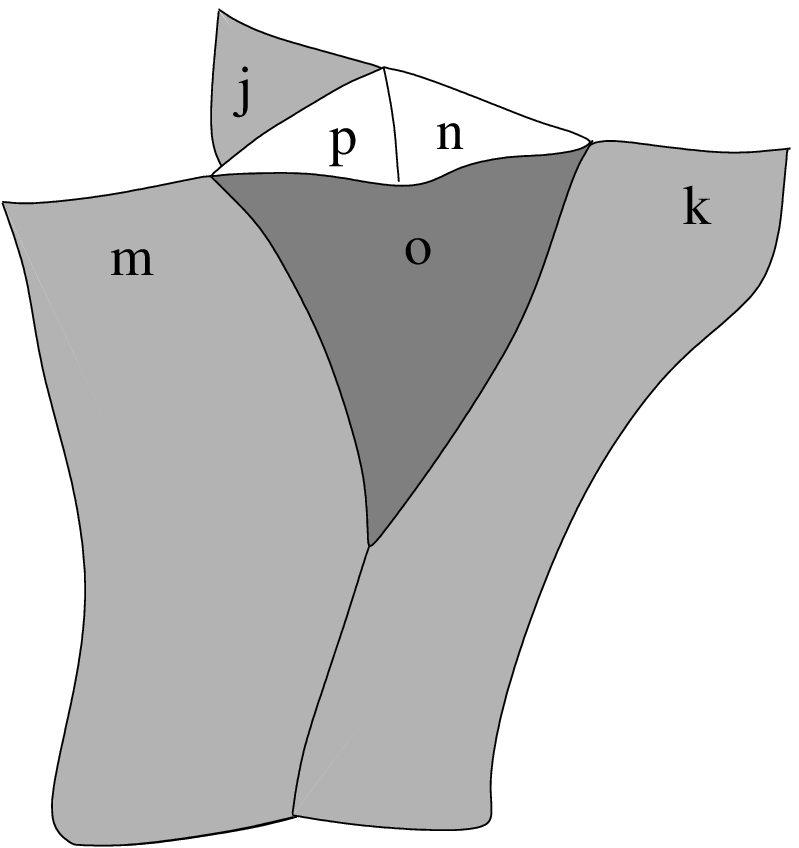}
\end{array}
\)} \caption{A set of discrete transitions representing one of the
contributing histories at a fixed value of the regulator. On the
right, the continuous {\em spin foam representation} when the
regulator is removed.} \label{vani}
\end{figure}

Each ${\rm Tr}[\stackrel{j_p}{\Pi}\!(W_{p})]$ in (\ref{final})
acts in $\Hk$ by creating a closed loop in the $j_{p}$ representation
at the boundary of the corresponding plaquette (Figs. \ref{pito} and
\ref{pitolon}).
Now, in order to obtain the {\em spin foam representation} we
introduce a non-physical ({\em coordinate time}) as follows: Instead
of working with one copy of the space manifold $\Sigma$ we consider $n_p(\epsilon)$ copies as  a
discrete folliation $\{\Sigma_p\}_{p=1}^{n_p(\epsilon)}$. Next we represent 
each of the ${\rm Tr}[\stackrel{j_p}{\Pi}\!(W_{p})]$ in (\ref{final})
on the corresponding $\Sigma_p$.
If one inserts the resolution of unity in
$\Hk$ between the slices, graphically \be \!\!\!\mathbbm{1}=\!\!\!\!\!\!\! \sum
\limits_{\gamma\subset \Sigma, \{j\}_{\gamma}} \!\!\!\!
|\gamma,\{j\}\rangle\langle\gamma,\{j\}|\begin{array}{c}
\includegraphics[width=5cm]{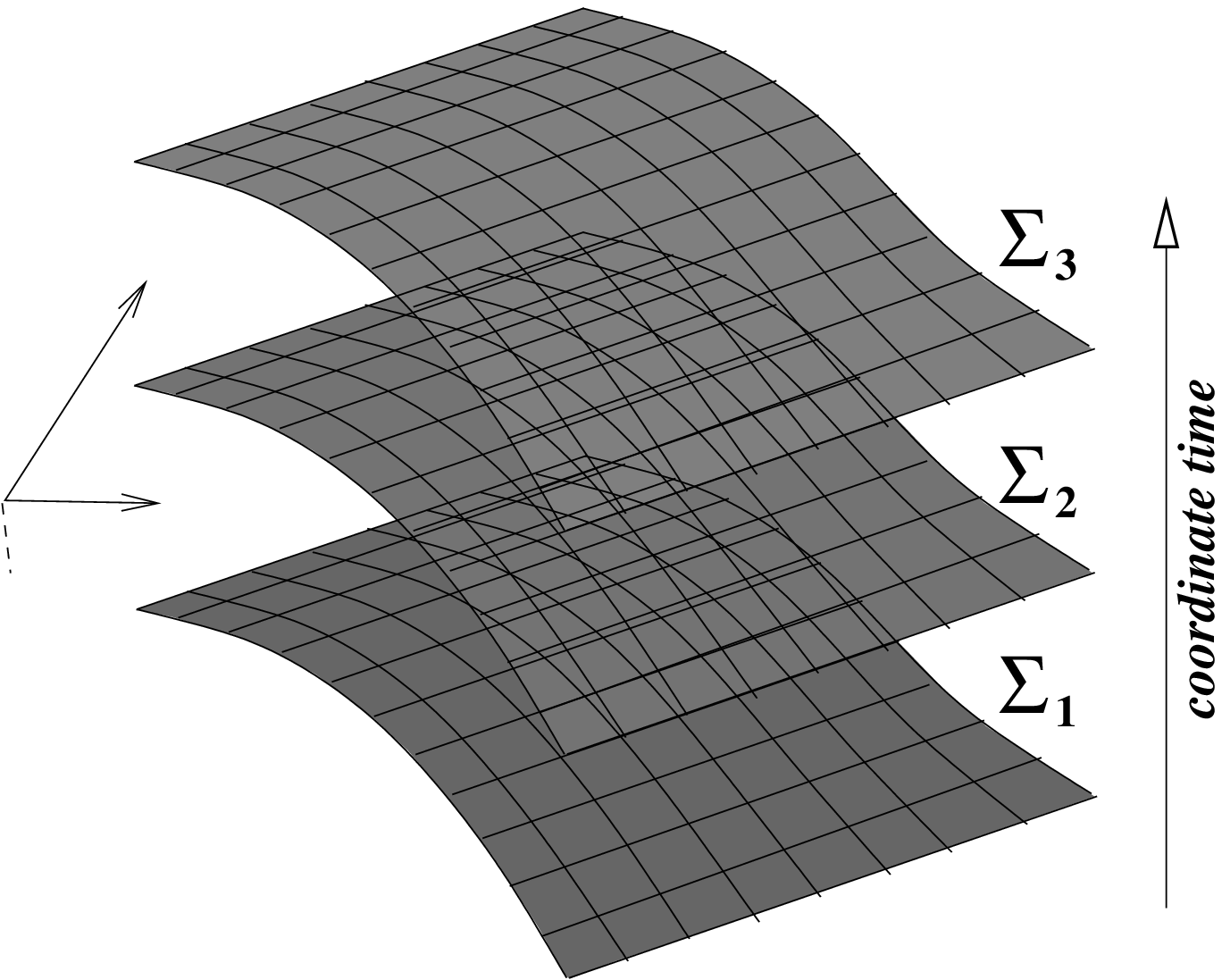}
                \end{array}
\label{e} \end{equation} where the sum is over the complete basis
of {\em spin network} states $\{|\gamma,\{j\}\rangle\}$---based on all
graphs $\gamma \subset \Sigma$ and with all possible spin
labelling---one arrives at a sum over
spin-network histories representation of $\langle s,s^{\prime}\rangle_p$. More
precisely, $\langle s^{\prime},s\rangle_p$ can be expressed as a sum over
amplitudes corresponding to a series of transitions that can be viewed
as the `time evolution' between the `initial' {\em spin network}
$s^{\prime}$ and the `final' {\em spin network} $s$. This is illustrated in
the two simple examples of Figs.
\ref{lupy} and \ref{vani}); on the r.h.s. we illustrate the continuum spin
foam picture obtained when the regulator is removed in the limit
$\epsilon\rightarrow 0$.

\begin{figure}[h]\!\!\!\!\!\!
\centerline{\hspace{0.5cm} \(
\begin{array}{c}
\includegraphics[height=4cm]{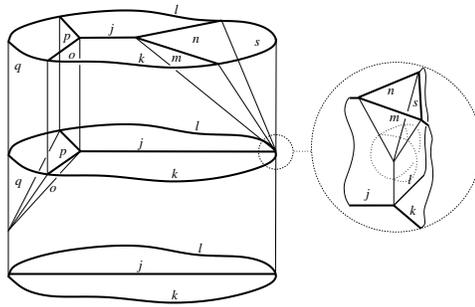}
\end{array}\) } \caption{A {\em spin foam} as the `colored' 2-complex
representing the transition between three different {\em spin
network} states. A transition vertex is magnified on the right.}
\label{spino}
\end{figure}

Spin network nodes evolve into edges while {\em spin network}
links evolve into 2-dimensional faces. Edges inherit the
intertwiners associated to the nodes and faces inherit the spins
associated to links. Therefore, the series of transitions can be
represented by a 2-complex whose 1-cells are labelled by
intertwiners and whose 2-cells are labelled by spins. The places
where the action of the plaquette loop operators create new links
(Figs. \ref{pitolon} and \ref{vani}) define  0-cells or
vertices. These foam-like structures are the so-called {\em spin
foams}. The {\em spin foam} amplitudes are purely combinatorial
and can be explicitly computed from the simple action of the loop
operator in $\Hk$.
The physical inner product takes the standard Ponzano-Regge form when
the {\em spin network}
states $s$ and $s^{\prime}$ have only 3-valent nodes. Explicitly,
\be \label{3dc} \langle s,s^{\prime}\rangle_p = \sum
\limits_{ F_{s\rightarrow s^{\prime}}} \ \prod_{f \subset F_{s\rightarrow s^{\prime}}} (2
j_f+1)^{\frac{\nu_f}{2}}
                \prod_{v\subset F_{s\rightarrow s^{\prime}}}
                \!\!\!\!\begin{array}{c}
                \includegraphics[width=2.3cm]{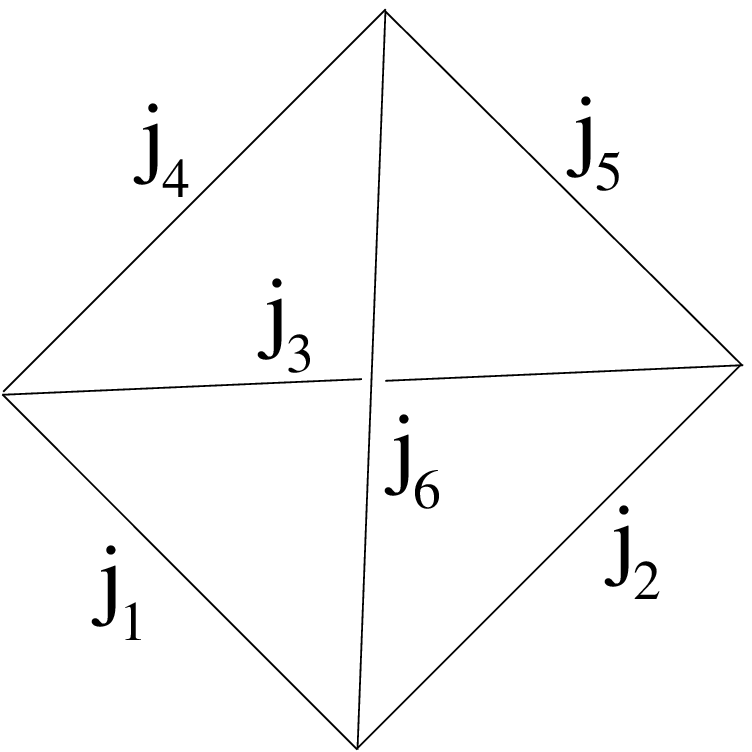}
                \end{array},
\end{equation} where the sum is over all the 
spin foams interpolating between $s$ and $s^{\prime}$ (denoted
$F_{s\rightarrow s^{\prime}}$, see Fig. \ref{spino}), $f\subset F_{s\rightarrow s^{\prime}}$ denotes
the faces of the spin foam (labeled by the spins $j_f$), $v\subset F_{s\rightarrow s^{\prime}}$
denotes vertices, and $\nu_f=0$ if $f
\cap s \not= 0 \wedge f \cap s^{\prime}\not= 0$, $\nu_f=1$ if $f
\cap s \not= 0 \vee f \cap s^{\prime}  \not= 0$,  and $\nu_f=2$ if
$f \cap s = 0 \wedge f \cap s^{\prime}= 0$. The tetrahedral
diagram denotes a $6j$-symbol: the amplitude obtained by means of
the natural contraction of the four intertwiners corresponding to
the 1-cells converging at a vertex.  More generally, for arbitrary
{\em spin networks}, the vertex amplitude corresponds to
$3nj$-symbols, and $\langle s,s^{\prime}\rangle_p$ takes the same general form.

\begin{figure}[h!!!!!!!!!!!!!]
 \centerline{\hspace{0.0cm}\(
\begin{array}{ccc}
\includegraphics[height=1cm]{lupo.eps} \\
\includegraphics[height=1cm]{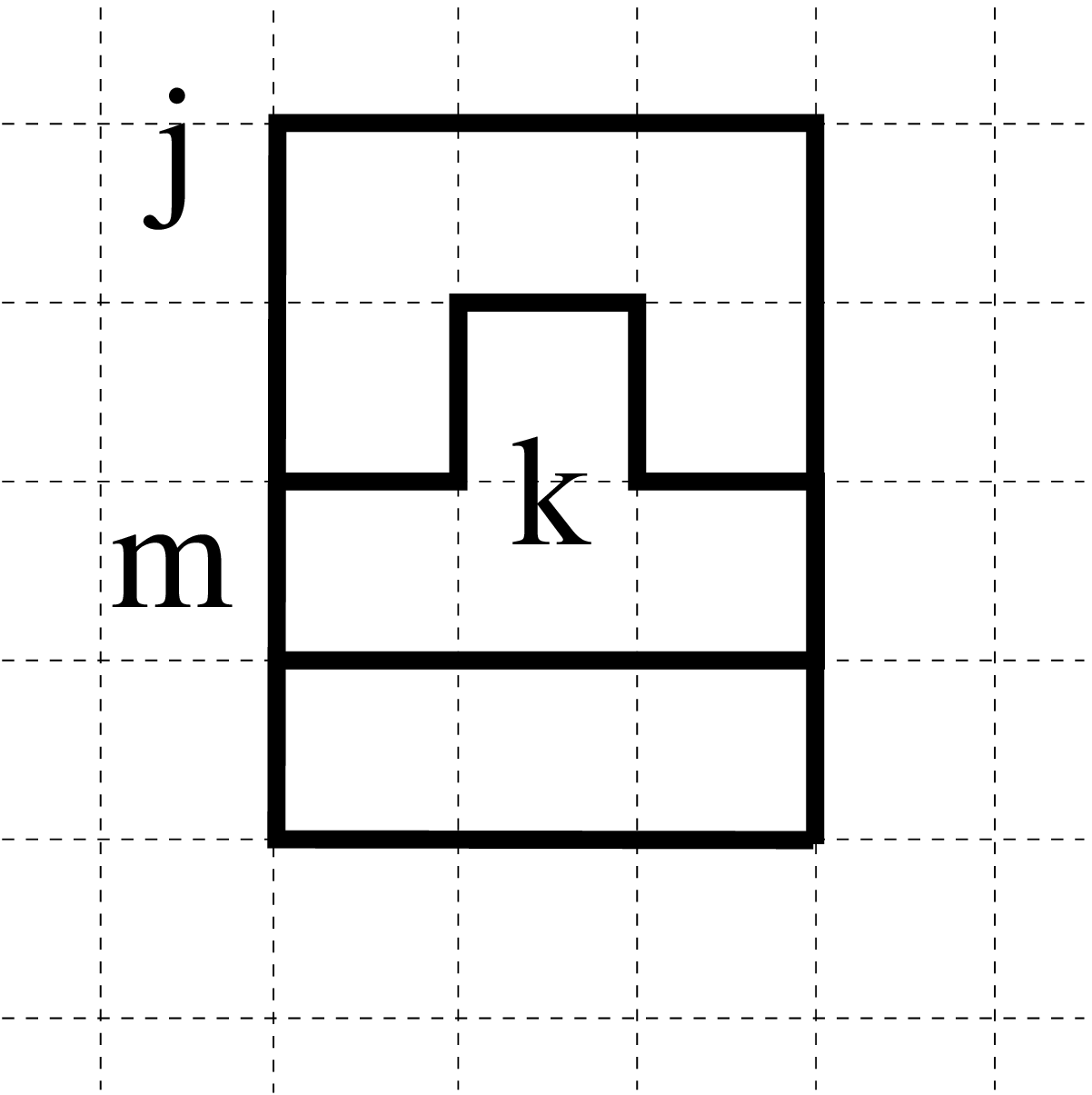}\\
\includegraphics[height=1cm]{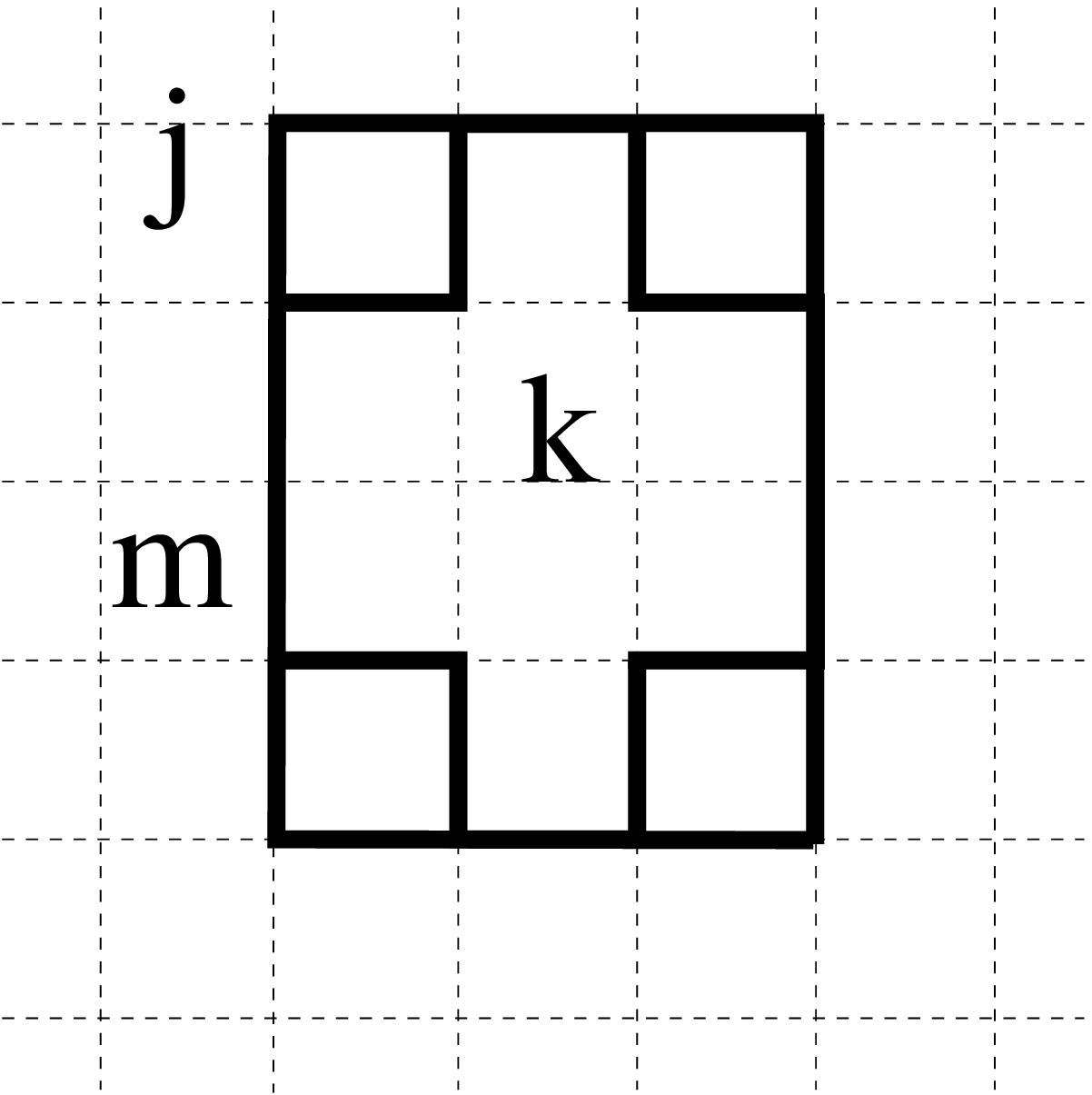}
\end{array}
\begin{array}{ccc}
\includegraphics[height=1cm]{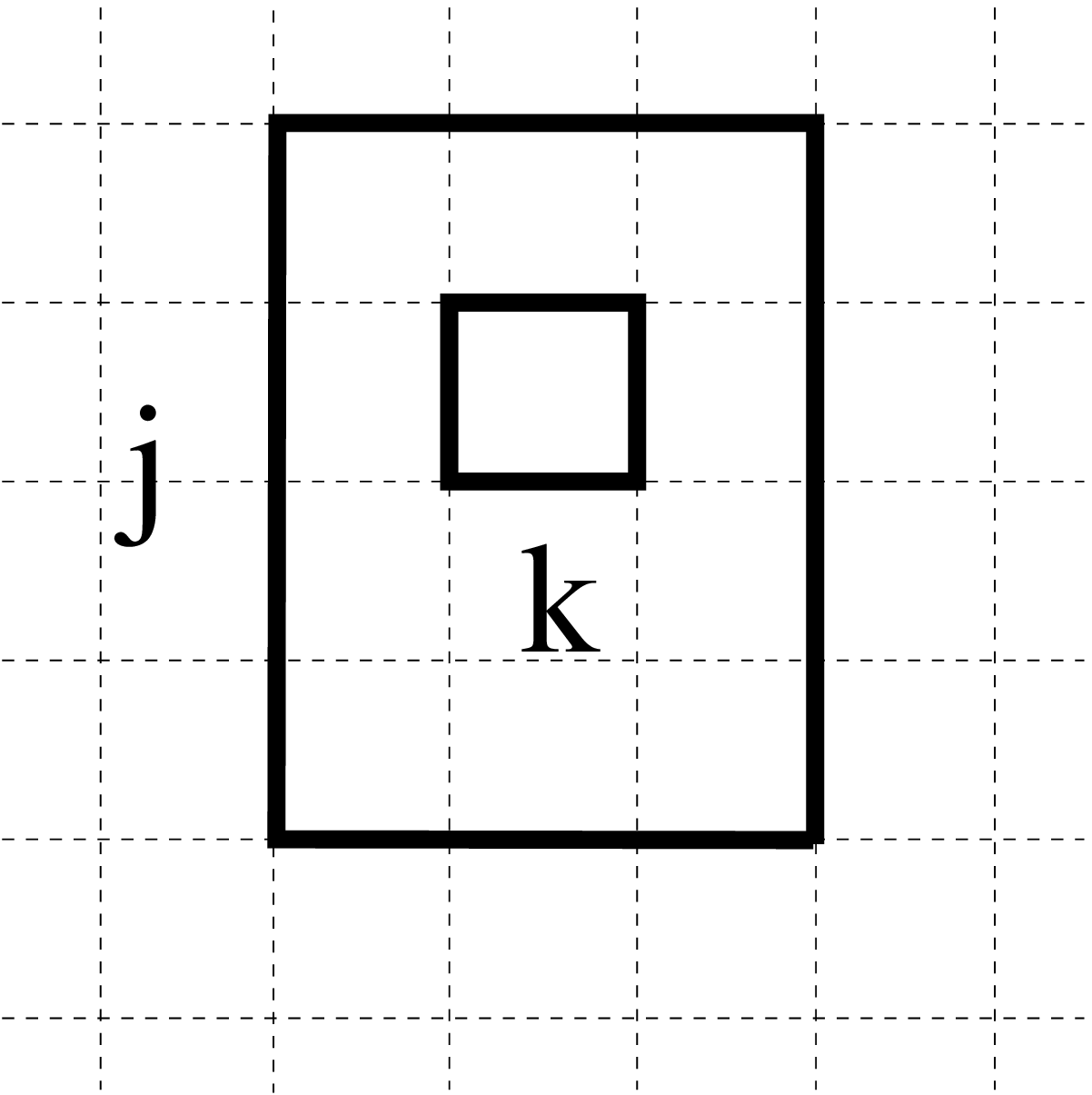} \\
\includegraphics[height=1cm]{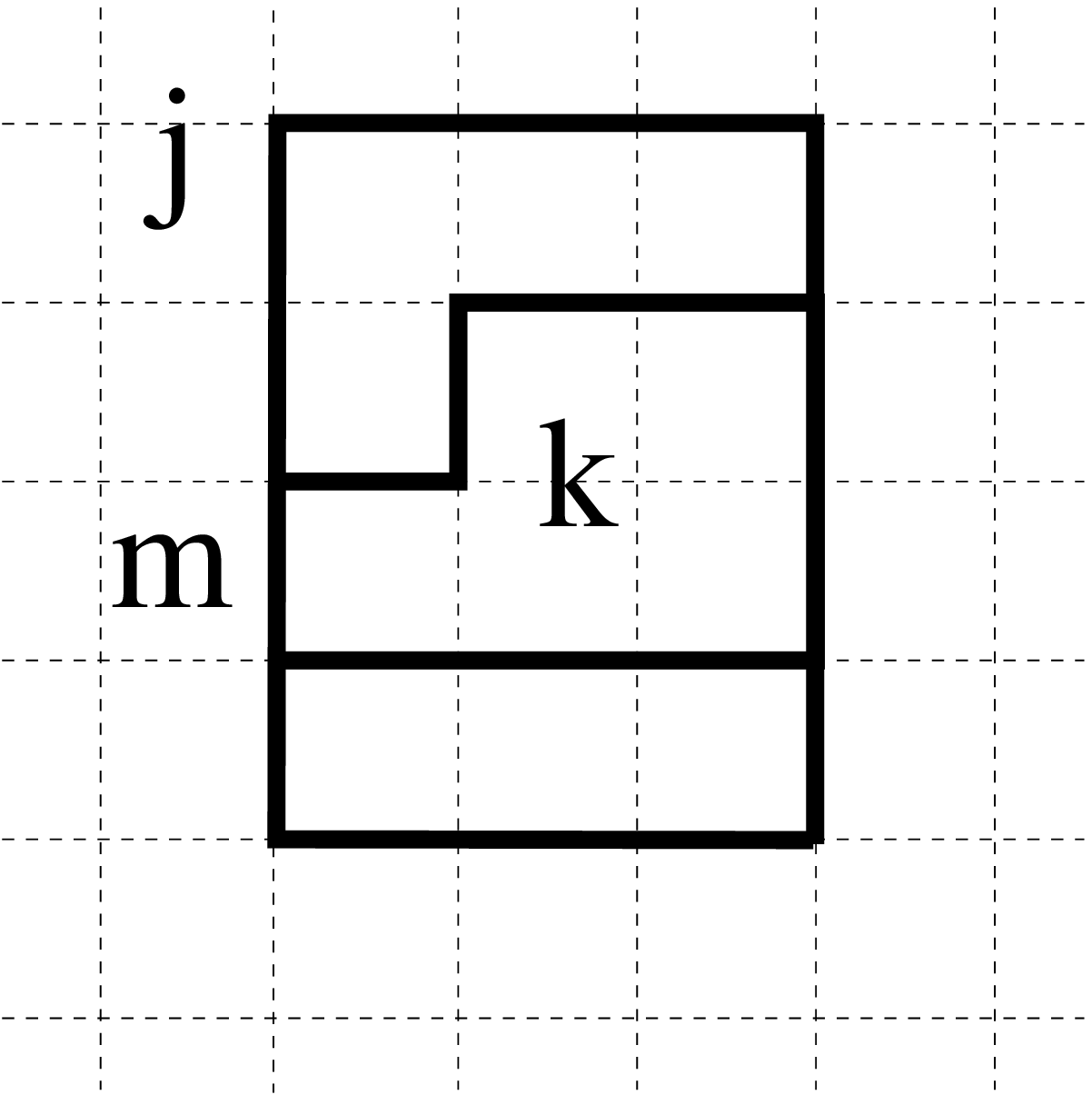}\\
\includegraphics[height=1cm]{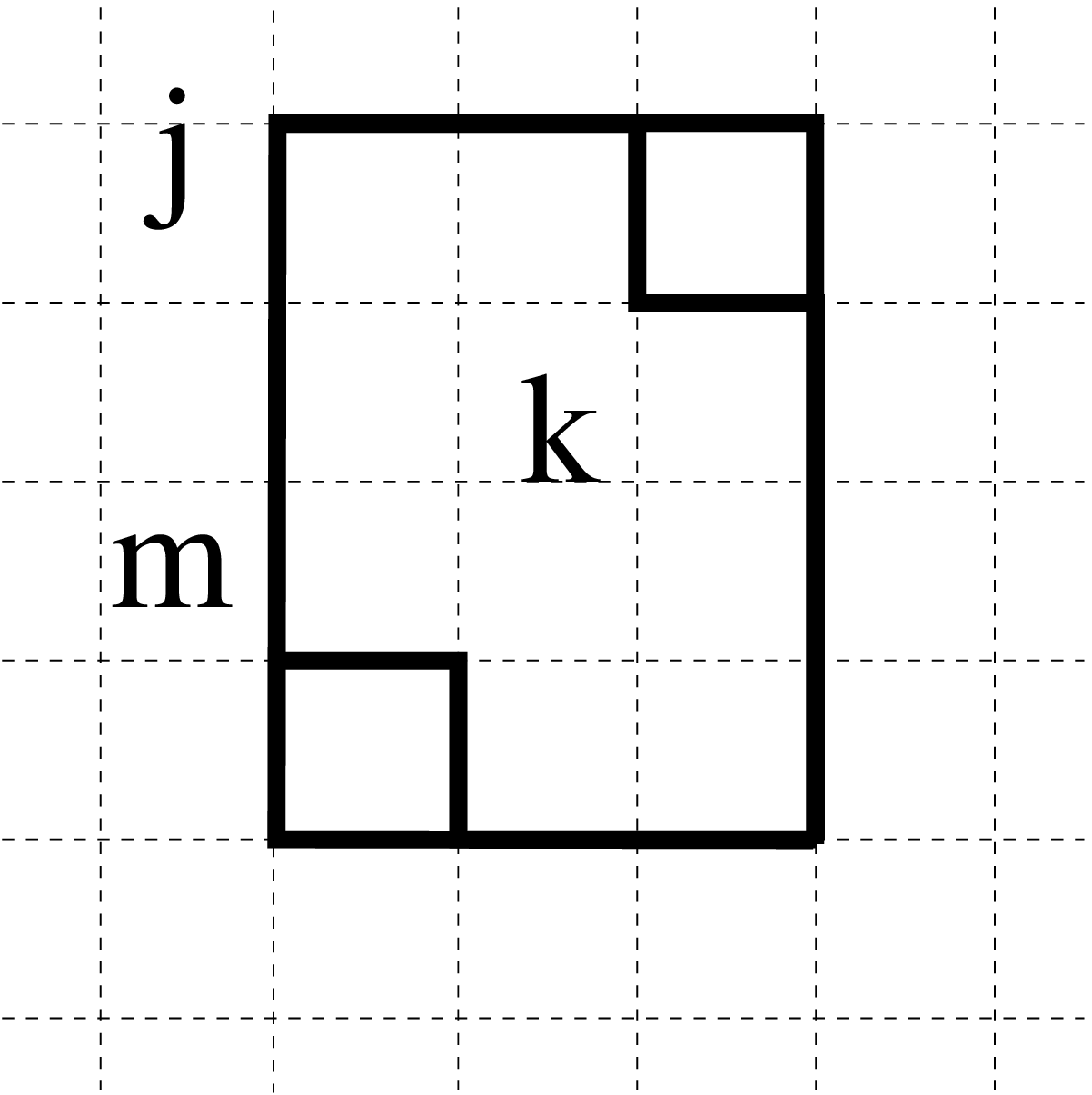}
\end{array}
\begin{array}{ccc}
\includegraphics[height=1cm]{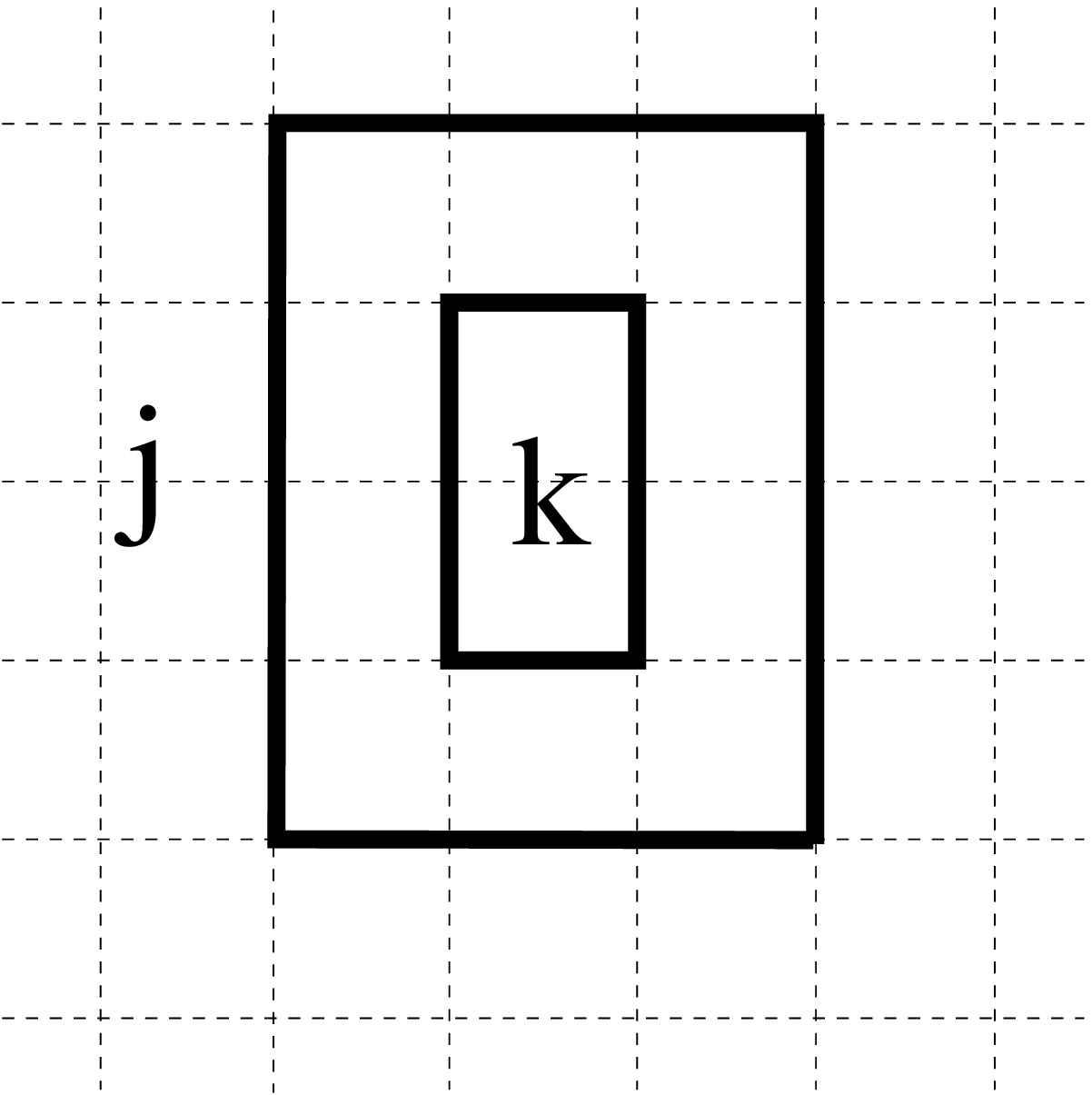} \\
\includegraphics[height=1cm]{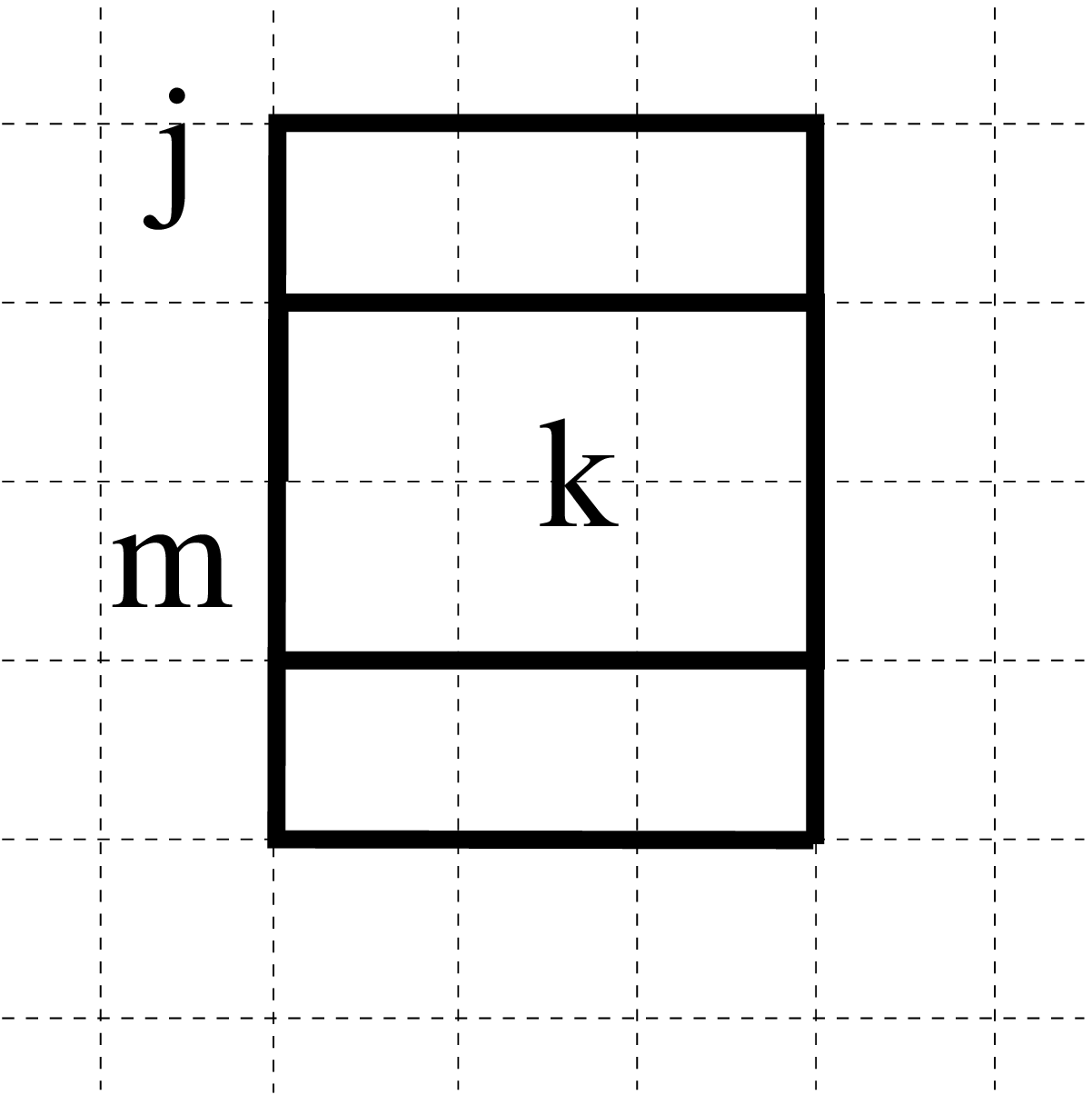}\\
\includegraphics[height=1cm]{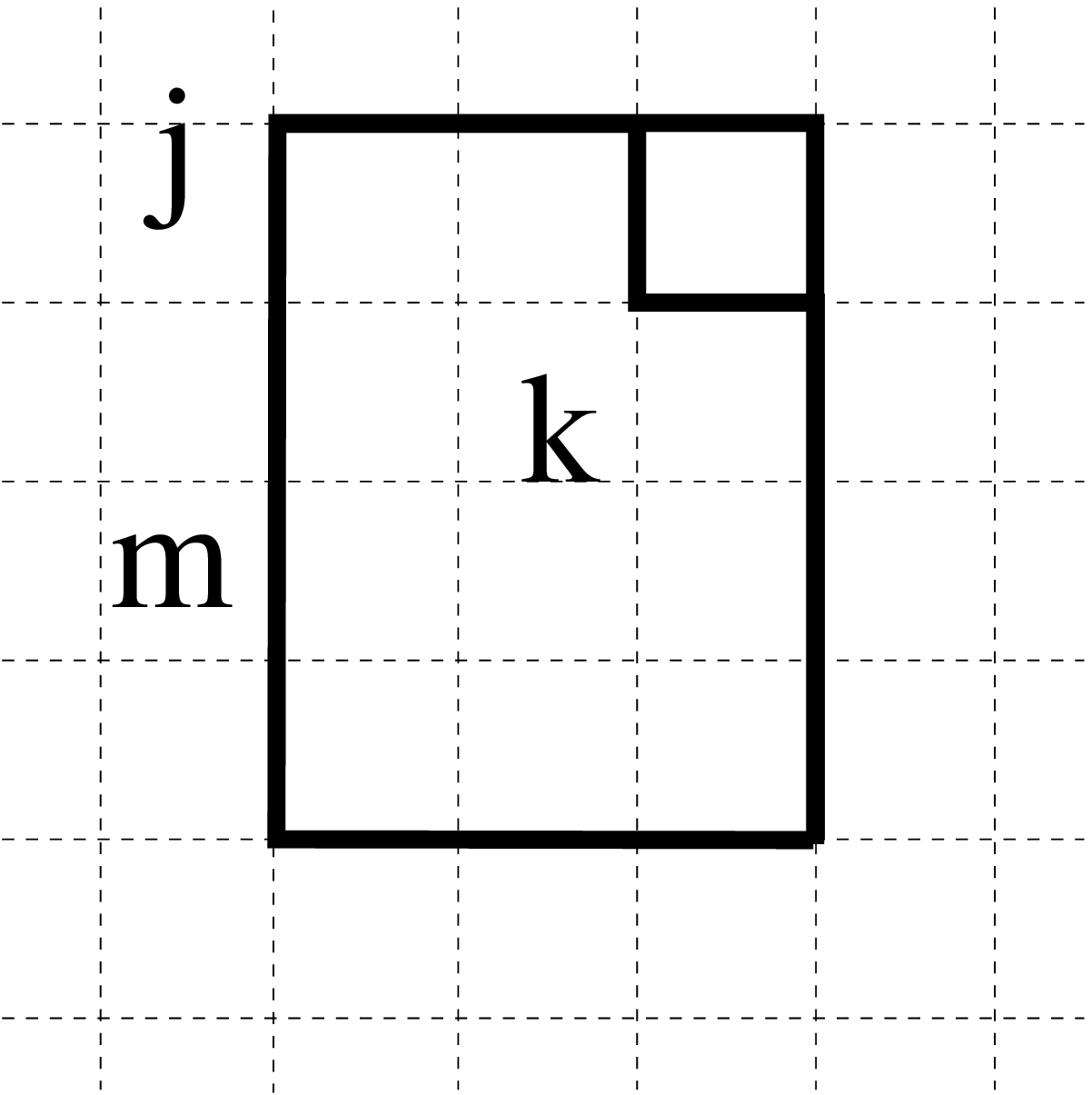}
\end{array}
\begin{array}{ccc}
\includegraphics[height=1cm]{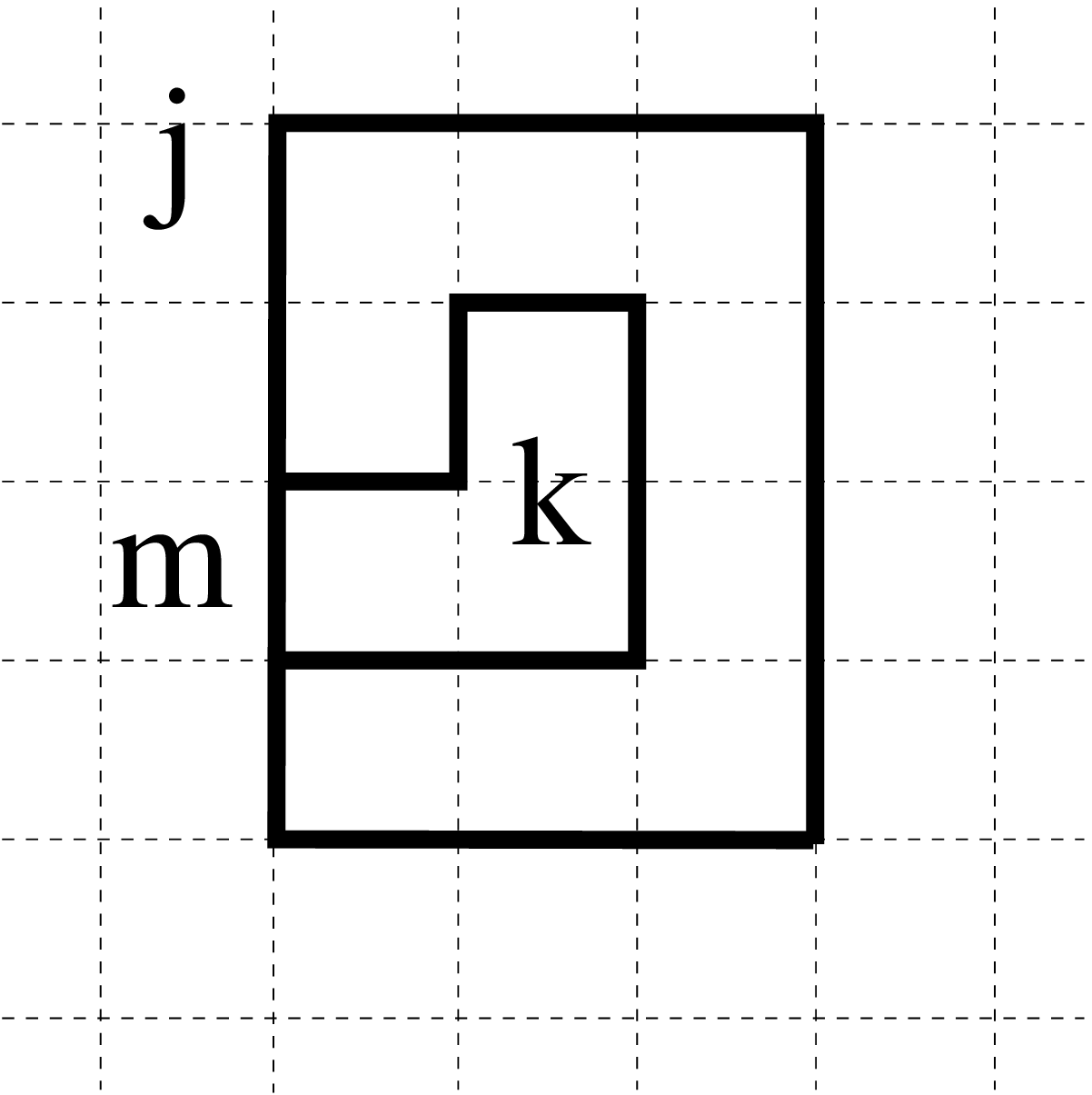} \\
\includegraphics[height=1cm]{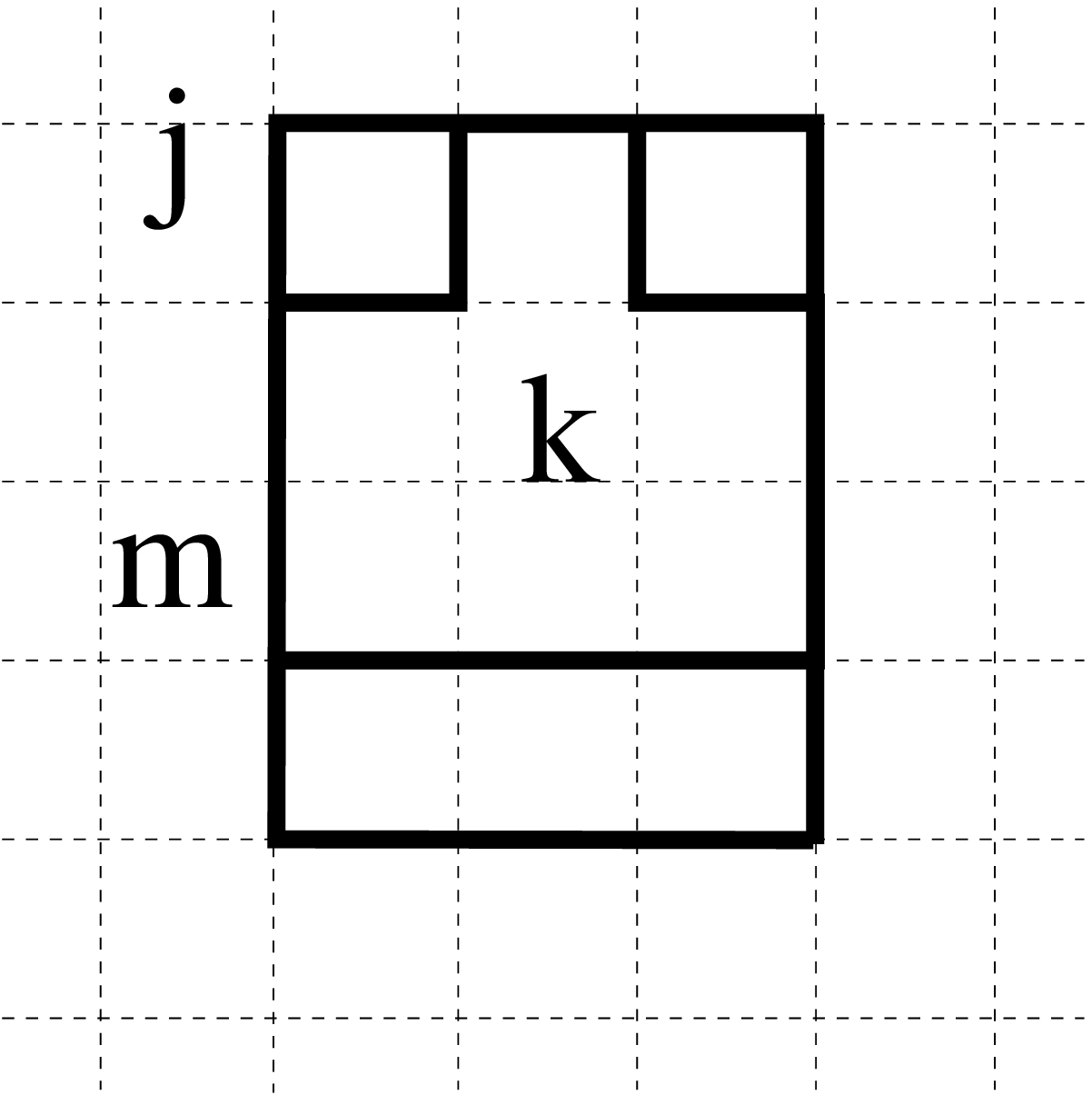}\\
\includegraphics[height=1cm]{lupo12.eps}
\end{array}\begin{array}{c}
\includegraphics[height=.3cm]{flecha.eps}
\end{array}\!\!\!
\begin{array}{c}
\includegraphics[height=3cm]{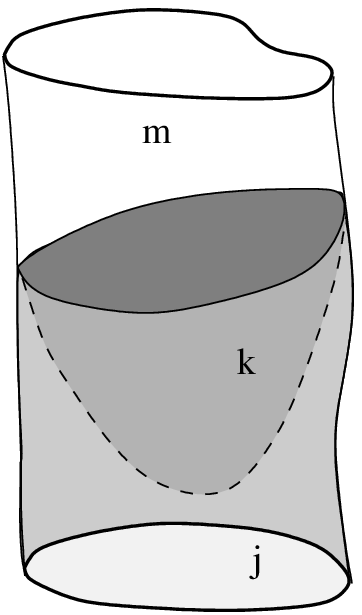}
\end{array}
\)} \caption{A different representation of the transition of figure \ref{lupy}. This spin-foam
is obtained by a different ordering choice in (\ref{final}).} \label{defy}
\end{figure}

Even though the ordering of the plaquette actions does not affect
the amplitudes, the {\em spin foam representation} of the terms in the
sum (\ref{3dc}) is highly dependent on that ordering. This is
represented in Fig~\ref{defy} where a spin foam equivalent to that
of Fig~\ref{lupy} is obtained by choosing an ordering of
plaquettes where those of the central region act first. One can
see this freedom of representation as an analogy of the gauge
freedom in the spacetime representation in the classical theory.

\begin{figure}[h!!!!]
\centerline{\hspace{0.5cm} \(\begin{array}{c}
\includegraphics[height=3.5cm]{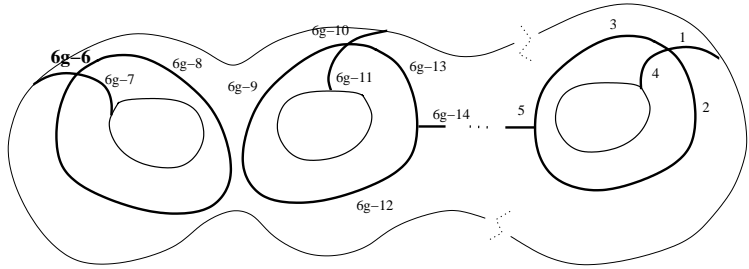}
\end{array}\) }
\caption{\small A spin-network basis of physical states for an arbitrary genus
$g$ Riemann surface. There are $6g-6$ spins labels (recall that 4-valent nodes carry an intertwiner quantum number).}
\label{toron}
\end{figure}

One can in fact explicitly construct a basis of $\Hp$ by choosing
an linearly independent set of representatives of the equivalence
classes defined in (\ref{null}). One of these basis is illustrated in
Fig.~\ref{toron}. The number of quantum numbers necessary to label the basis
element is $6g-6$ corresponding to the dimension of the moduli space of $SU(2)$ flat
connections on a Riemann surface of genus $g$. This is the number of degrees
of freedom of the classical theory. In this way we
arrive at a fully combinatorial definition of the standard $\Hp$
by reducing the infinite degrees of freedom of the kinematical
phase space to finitely many by the action of the generalized
projection operator $P$.

\subsection{Quantum spacetime as gauge-histories}\label{gaugy}

What is the geometric meaning of the spin foam configurations? Can we
identify the spin foams with {\em ``quantum spacetime
configurations''}?  The answer to the above questions is, strictly
speaking, in the negative in agreement with our discussion at the end of
Sec.~\ref{valin}. This conclusion can be best illustrated by looking
first at the simple example in 2+1 gravity where $M=S^2\times \R$
($g=0$).  In this case the spin foam configurations appearing in the
transition amplitudes look locally the same to those appearing in the
representation of $P$ for any other topology.  However, a close look
at the physical inner product defined by $P$ permits to conclude that
the physical Hilbert space is one dimensional---the classical theory
has zero degree of freedom and so there is no non-trivial Dirac
observable in the quantum theory. This means that the sum over spin foams
in (\ref{3dc}) is nothing else but a sum over {\em pure gauge degrees
of freedom} and hence no physical interpretation can be associated to
it. The spins labelling the intermediate spin foams do not correspond
to any measurable quantity. For any other topology this still holds
true, the true degrees of freedom being of a global topological
character. This means that in general (even when local excitations are
present as in 4d) the spacetime geometric interpretation of the spin
foam configurations is subtle.  This is an important point that is
often overlooked in the literature: one cannot interpret the spin foam
sum of (\ref{3dc}) as a sum over geometries in any obvious way. Its
true meaning instead comes from the averaging over the gauge orbits
generated by the quantum constraints that defines $P$---recall the
classical picture Fig.~\ref{phase}, the discussion around
eq. (\ref{pipi}), and the concrete implementation in 2+1 where $U(N)$
in (\ref{exists}) is the unitary transformation representing the
orbits generated by $F$. Spin foams represent a {\em gauge history} of
a kinematical state. A sum over gauge histories is what defines $P$ as
a means for extracting the true degrees of freedom from those which
are encoded in the kinematical boundary states.

Here we studied interpretation of the {\em spin foam representation} in the precise context
of our toy example; however, the validity of the conclusion is of general
character and holds true in the case of physical interest four dimensional
LQG.  Although, the quantum numbers labelling the spin foam configurations
correspond to eigenvalues of {\em kinematical} geometric quantities such as
length (in 2+1) or area (in 3+1) LQG, their physical meaning and {\em
measurability} depend on dynamical considerations (for instance the naive
interpretation of the spins in 2+1 gravity as quanta of physical length is
shown here to be of no physical relevance).  Quantitative notions such as time, or
distance as well as qualitative statements about causal structure or time
ordering are misleading (at best) if they are naively constructed in
terms of notions arising from an interpretation of {\em spin foams} as quantum
spacetime configurations\footnote{The discussion of this
section is a direct consequence of Dirac's perspective applied to the {\em spin foam
representation}.}.

\section{{\em Spin foam models} in four dimensions}

We have studied 2+1 gravity in order to introduce the qualitative features of
the {\em spin foam representation} in a precise setting. Now we discuss some
of the ideas that are pursued for the physical case of 3+1 LQG.

\subsubsection*{Spin foam representation of canonical LQG}

There is no complete construction of the physical inner product of LQG
in four dimensions. The {\em spin foam representation} as a device for
its definition was originally introduced in the canonical formulation
by Rovelli (see Rovelli C. (2005)). In 4-dimensional LQG difficulties
in understanding dynamics are centered around understanding the space
of solutions of the quantum scalar constraint $\hat S$.
The physical inner
product formally becomes
\begin{eqnarray} \label{vanino}&&\!\!\!\! \!\!\!\left\langle Ps,
s^{\prime}\right\rangle_{\vani \rm diff}=
\int {D}[N] \sum \limits^{\infty}_{n=0} \frac{i^{n}}{n!}\langle\left[\int
\limits_{\Sigma} N(x) \widehat {S}(x)\right]^n \!\!\!\! s,
s^{\prime}\rangle_{\vani\rm diff},
\end{eqnarray}
\vskip-.1cm \noindent where $\langle\ ,\ \rangle_{\vani \rm diff}$ denotes the
inner product in the Hilbert space of diff-invariant states, and
the exponential in (the field theoretical analog of) (\ref{pipi}) has been expanded in powers.

\begin{figure}[h]
\centerline{\hspace{1.5cm} \(\!\!\!\!\!\! \!\!\!\!\!\!\!\!\!\!\!\! \!\!\!\!\!\! \int
\limits_{\Sigma} N(x) \widehat {S}(x) \ \rhd
\!\!\!\!\!\!\!\!\!\!\!\!
\begin{array}{c}
\includegraphics[height=2cm]{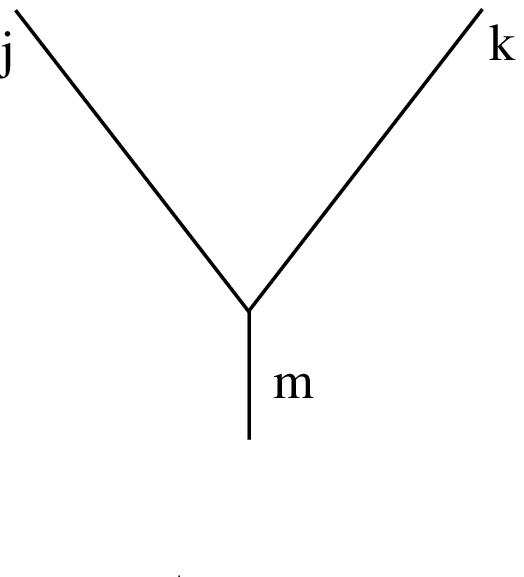}
\end{array} \!\!\!\!\!\!\!\!\!\!\!\! =\sum \limits_{nop} N(x_n) S_{nop} \!\!\!\!\!\!\!\!
\begin{array}{c}
\includegraphics[height=2cm]{tero3.eps}
\end{array}
\begin{array}{c}
\includegraphics[height=.3cm]{flecha.eps}
\end{array}
\begin{array}{c}
\includegraphics[height=2cm]{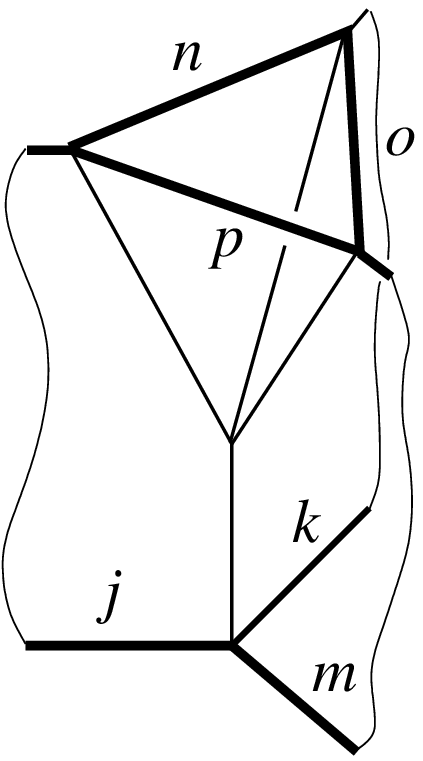}
\end{array}
\) } \caption{The action of the scalar constraint and its {\em
spin foam representation}. $N(x_n)$ is the value of $N$ at the
node and $S_{nop}$ are the matrix elements of $\widehat S$. }
\label{actionH}
\end{figure}

Smooth loop states are naturally annihilated by $\widehat {S}$
(independently of any quantization ambiguity (Jacobson T. \& Smolin
L. (1988) and Smolin L. \&
 Rovelli, C. (1990))).  Consequently, $\widehat S$ acts only on {\em spin
network} nodes.  Generically, it does so by creating new links and
nodes modifying the underlying graph of the {\em spin network} states
(Figure \ref{actionH}).

In a way that is qualitatively similar to what we found in the
concrete implementation of the curvature constraint in 2+1
gravity, each term in the sum (\ref{vanino}) represents a series of
transitions---given by the local action of $\widehat {S}$ at {\em
spin network} nodes---through different {\em spin network} states
interpolating the boundary states $s$ and $s^{\prime}$
respectively. The action of $\So$ can be visualized as an
`interaction vertex' in the `time' evolution of the node (Figure
\ref{actionH}). As in 2+1 dimensions, equation (\ref{vanino}) can
be pictured as sum over `histories' of {\em spin networks}
pictured as a system of branching surfaces described by a
2-complex whose elements inherit the representation labels on the
intermediate states (see Fig.~\ref{spino}). The value of the
`transition' amplitudes is controlled by the matrix elements of
$\widehat S$.

\subsubsection*{Spin foam representation in the Master Constraint Program}

The previous discussion is formal. One runs into technical
difficulties if one tries to implement the construction of the 2+1
gravity in this case. The main reason for this is the fact that
the constrain algebra does not close with structure constants in
the case of 3+1 gravity\footnote{In 2+1 gravity the constraint
algebra correspond to the Lie algebra of $ISO(3)$ (isometries of
Euclidean flat spacetime). There are no local degrees of freedom
and the underlying gauge symmetry has a non dynamical structure.
In 3+1 gravity the presence of gravitons changes that. The fact
that the constraint algebra closes with structure functions means
that the gauge symmetry structure is dynamical or field dependent.
This is the key difficulty in translating the simple results of
2+1 into 3+1 dimensions.}. In order to circumvent this problem
Thiemann recently proposed  to impose one
single {\em master} constraint defined as
\begin{equation}
M= \int \limits_{\Sigma} dx^3 \ \frac{S^2(x)- q^{ab} V_{a}(x)
V_{b}(x)}{\sqrt{{\rm det}\ q(x)}},
\end{equation}
where $q^{ab}$ is the space metric and $V_a(x)$ is the vector constraint.
Using techniques developed by Thiemann this constraint can indeed
be promoted to a quantum operator acting on $\Hk$. The physical
inner product could then be defined as
\begin{equation}
\langle s,s^{\prime}\rangle_p:=\lim_{T\rightarrow \infty} \langle s,\int
\limits_{-T}^{T}\ dt \ e^{i t \widehat M}s^{\prime}\rangle.
\end{equation}
A spin-foam-representation of the previous expression is obtained by
splitting the $t$-parameter in discrete steps and writing
\begin{equation} e^{it \widehat M}=\lim_{n\rightarrow \infty}\ [e^{it
\widehat M/n}]^n =\lim_{n\rightarrow \infty}\ [1+it \widehat
M/n]^n.
\end{equation}
The {\em spin foam representation} follows from the fact that the action of the
basic operator $1+it \widehat M/n$ on a {\em spin network}  can be
written as a linear combination of new {\em spin networks}  whose
graphs and labels have been modified by the creation of new nodes
(in a way qualitatively analogous to the local action shown in
Figure \ref{actionH}). An explicit derivation of the physical
inner product of 4d LQG along these lines is under current
investigation.

\subsubsection*{Spin foam representation: the covariant perspective}

In four dimensions the {\em spin foam representation} of LQG has also
been motivated by lattice discretizations of the path integral of
gravity in the covariant formulation (for recent reviews see Perez
A. (2003) and Oriti D. (2001)).  In four dimensions
there are two main lines of approach; both are based on classical
formulations of gravity based on modifications of the BF-theory
action.

The first approach is best represented by the \BC model (Barrett
J.W. \& Crane L. (1998)) and corresponds to the quantization attempt
of the classical formulation of gravity based on the Plebanski action
\begin{equation}\label{pleb}
S[B,A,\lambda]=\int {\rm Tr}\left[B\wedge F(A) +
\lambda \ B \wedge B\right],
\end{equation}
\vskip-.1cm \noindent where $B$ is an $so(3,1)$-valued two-form
$\lambda$ is a Lagrange multiplier imposing a quadratic constraint on
the $B$'s whose solutions include the sector $B={\star}(e\wedge e)$,
for a tetrad $e$, corresponding to gravity in the tetrad
formulation. The key idea in the definition of the model is that the
path integral for BF-theory, whose action is $S[B,A,0]$, \vskip-.3cm
\be P_{topo}=\int \ D[B] D[A] \ {\rm exp}\left[i \int {\rm
Tr}\left[B\wedge F \right]\right]\label{Topo}\end{equation}
\vskip-.1cm \noindent can be defined in terms of {\em spin foams} by a simple generalization of the
construction of Sec.~\ref{pipo}. Notice that the formal structure of the
action $S[B,A,0]$ is analogous to that of the action of 2+1 gravity
(\ref{bfaction}) (see Baez J.C. (2000)). The \BC model aims at providing a
definition of the path integral of gravity formally written as\vskip-.3cm
\begin{equation}\label{heu}
P_{\va GR}=\int \ D[B] D[A] \ \delta\left[B\rightarrow \star(
e\wedge e)\right] \  {\rm exp}\left[i \int {\rm Tr} \left[B\wedge
F\right]\right],\end{equation} \vskip-.1cm \noindent
 where the measure $D[B] D[A] \delta[B\rightarrow \star(
e\wedge e)]$ restricts the sum in (\ref{Topo})
to those configurations of the topological theory satisfying the
constraints $B={\star}( e \wedge e)$ for some tetrad $e$. 
The remarkable fact is that the constraint $B={\star}( e \wedge e)$ can be
directly implemented on the {\em spin foam} configurations of
$P_{topo}$ by appropriate restriction on the allowed spin labels and
intertwiners. All this is possible if a regularization is provided, consisting of a
cellular decomposition of the spacetime manifold. The
key open issue is however how to get rid of this regulator. A
proposal for a regulator independent definition is that of
the {\em group field theory formulation} (Oriti D. (2005)).

A second proposal is the one recently introduced by Freidel and Starodubtsev
2005 based on the formulation McDowell-Mansouri action of Riemannian
gravity given by \be S[B,A]=\int {\rm Tr}[B\wedge F(A)-\frac{\alpha}{4}
B\wedge B \gamma_5], \ee where $B$ is an $so(5)$-valued two-form, $A$ an
$so(5)$ connection, $\alpha= G\Lambda/3\approx 10^{-120}$ a coupling constant,
and the $\gamma_5$ in the last term produces the symmetry braking
$SO(5)\rightarrow SO(4)$. The idea is to define $P_{\va GR}$ as a power series
in $\alpha$, namely \be\label{ssee}\nonumber P_{\va
GR}=\sum\limits^{\infty}_{n=0} \frac{(-i\alpha)^n}{4^n n!} \int D[B]D[A] ({\rm
Tr}[B\wedge B \gamma_5])^n\ {\rm exp}\left[i\int {\rm Tr}[B\wedge F]
\right]. \ee Notice that each term in the sum is the expectation value of a
certain power of $B$'s in the well understood topological BF field theory. A
regulator in the form of a cellular decomposition of the spacetime manifold
is necessary to give a meaning to the former expression. Due to the absence of
local degrees of freedom of BF-theory it is expected that the regulator can be
removed in analogy to the 2+1 gravity case. It is important to show
that removing the regulator does not produce an uncontrollable set of
ambiguities (see remarks below regarding {\em renormalizability}).

\subsection{The UV problem in the background independent context}\label{amby}

In the {\em spin foam representation}, the functional integral for
gravity is replaced by a sum over amplitudes of combinatorial objects
given by foam-like configurations ({\em spin foams}). This is a direct
consequence of the background independent treatment of the
gravitational field degrees of freedom. As a result there is no place
for the UV divergences that plague standard quantum field theory. The
combinatorial nature of the fundamental degrees of freedom of geometry
appears as a regulator of all the interactions. This seem to be a
common feature of all the formulations referred to in this
chapter. Does it mean that the UV problem in LQG is resolved? The
answer to this question remains open for the following reason. All the
definitions of spin foams models require the introduction of some kind
of regulator generically represented by a space (e.g., in the
canonical formulation of 2+1 gravity or in the master-constraint
program) or spacetime lattice (e.g. in the \BC model or in the
Freidel-Starodubtsev prescription). This lattice plays a role of a UV
regulator in more or less the same sense as a UV cut-off ($\Lambda$)
in standard QFT. The UV problem in standard QFT is often associated to
divergences in the amplitudes when the limit $\Lambda\rightarrow
\infty$ is taken. The standard renormalization procedure consists of
taking that limit while appropriately tuning the bare parameters of
the theory so that UV divergences cancel to give a finite
answer. Associated to this process there is an intrinsic ambiguity as
to what values certain amplitudes should take. These must be fixed by
appropriate comparison with experiments (renormalization
conditions). If only a finite number or renormalization conditions are
required the theory is said to be {\em renormalizable}. The ambiguity
of the process of removing the regulator is an intrinsic feature of
QFT.

The background independent treatment of gravity in LQG or the {\em spin
foam models} we have described here do not escape to this general
considerations (Perez A. (2005)). Therefore, even though no UV divergences can arise as
a consequence of the combinatorial structure of the gravitational
field, the heart of the UV problem is now to be found in the
potential ambiguities associated with the elimination of the
regulator. This remains an open problem for all the
attempts of quantization of gravity in 3+1 dimensions. The problem
takes the following form in each of the approaches presented in
this chapter:
\begin{itemize}
\item The removal of the regulator in the 2+1 case is free of
ambiguities and hence free of any UV problem (Perez A. (2005)).

\item In the case of the master constraint program one can
explicitly show that there is a large degree of ambiguity
associated to the regularization procedure (Perez A. (2005)). It remains to be shown
whether this ambiguity is reduced or disappears when the
regulators are removed in the definition of $P$.

\item The \BC model is discretization dependent. No clear-cut
prescription for the elimination of the triangulation dependence is known.

\item The Freidel-Starodubtsev prescription suffers (in principle) from the
ambiguities associated to the definition of the expectation value
of the B-monomials appearing in (\ref{ssee}) before the regulator is
removed\footnote{There are various prescriptions in the literature
on how to define this monomials. They are basically constructed in
terms of the insertion of appropriate sources to construct a BF
generating function. All of them are intrinsically ambiguous, and
the degree of ambiguity grows with the order of the monomial. The
main source of ambiguity resides in the issue of where in the
discrete lattice to act with the functional source derivatives.}.
It is hoped that the close relationship with a topological theory might cure
these ambiguities although this remains to be shown.

\end{itemize}
Progress in the resolution of this issue in any of these
approaches would represent a major breakthrough in LQG.

\section*{Acknowledgements}
I would like to thank M. Mondragon and A. Mustatea for careful reading
of the manuscript and D. Oriti for his effort and support in this
project.

\end{document}